\documentstyle[pra,aps,multicol,epsfig]{revtex}

\input epsf.tex

\newcommand{\be}{\begin{equation}}
\newcommand{\ee}{\end{equation}}
\newcommand{\ba}{\begin{eqnarray}}
\newcommand{\ea}{\end{eqnarray}}
\newcommand{\ban}{\begin{eqnarray*}}
\newcommand{\ean}{\end{eqnarray*}}

\newcommand{\ket}[1]{\mbox{$ | #1 \rangle $}}
\newcommand{\bra}[1]{\mbox{$ \langle #1 | $}}

\newcommand{\demi}{\frac{1}{2}}

\newcommand{\real}{\begin{picture}(8,8)\put(0,0){R}\put(0,0){\line(0,1){7}}\end{picture}}

\newcommand{\one}{\leavevmode\hbox{\small1\normalsize\kern-.33em1}}

\def\tr{\mbox{tr}}

\begin{document}

\title{Secrecy extraction from no-signalling correlations}
\author{Valerio Scarani, Nicolas Gisin, Nicolas Brunner}
\address{
Group of Applied Physics, University of Geneva, 20, rue de
l'Ecole-de-M\'edecine, CH-1211 Geneva 4, Switzerland}

\author{Lluis Masanes}
\address{School of Mathematics, University of Bristol, Bristol BS8 1TW, United Kingdom}

\author{Sergi Pino, Antonio Ac\'{\i}n}
\address{
ICFO - Institut de Ci\`encies Fot\`oniques, Mediterranean
Technology Park, 08860 Castelldefels (Barcelona), Spain}

\date{\today}
\maketitle

\begin{abstract}

Quantum cryptography shows that one can guarantee the secrecy of
correlation on the sole basis of the laws of physics, that is
without limiting the computational power of the eavesdropper. The
usual security proofs suppose that the authorized partners, Alice
and Bob, have a perfect knowledge and control of their quantum
systems and devices; for instance, they must be sure that the
logical bits have been encoded in true qubits, and not in
higher-dimensional systems. In this paper, we present an approach
that circumvents this strong assumption. We define protocols, both
for the case of bits and for generic $d$-dimensional outcomes, in
which the security is guaranteed by the very structure of the
Alice-Bob correlations, under the no-signalling condition. The
idea is that, if the correlations cannot be produced by shared
randomness, then Eve has poor knowledge of Alice's and Bob's
symbols. The present study assumes, on the one hand that the
eavesdropper Eve performs only individual attacks (this is a
limitation to be removed in further work), on the other hand that
Eve can distribute any correlation compatible with the
no-signalling condition (in this sense her power is greater than
what quantum physics allows). Under these assumptions, we prove
that the protocols defined here allow extracting secrecy from
noisy correlations, when these correlations violate a Bell-type
inequality by a sufficiently large amount. The region, in which
secrecy extraction is possible, extends within the region of
correlations achievable by measurements on entangled quantum
states.

\end{abstract}


\begin{multicols}{2}

\section{Introduction}

Quantum physics has been shown to provide a means to distribute
correlations at a distance, whose secrecy can be guaranteed by the
laws of physics, without any assumption on the computational power
of the eavesdropper. This is the nowadays largely studied field of
quantum cryptography (or quantum key distribution, QKD), the most
mature development of quantum information science \cite{review1}.
The fact itself, that quantum physics can be used to distribute
secrecy, is safe: if the authorized partners share a maximally
entangled state, then secrecy is definitely guaranteed. But of
course, one must verify that secrecy is not immediately spoiled by
any small departure from this ideal case; this is why much
theoretical research has been devoted to the derivation of
rigorous bounds for the security of quantum cryptography
\cite{review2}. Still, a lot of questions remain unsolved: for
instance, the theorists, who find security proofs, and the
experimentalists, who realize devices, tend to make different and
often incompatible assumptions when figuring their schemes out.

In particular, an assumption in theoretical proofs has gone
unnoticed until recently \cite{bhk,prl}: one assumes that the
logical bits are encoded in quantum systems whose dimension is
under perfect control (generally, qubits). Why do we question this
assumption? First, because it is interesting in itself to ask,
whether one can remove an assumption, that is, whether one can
base the studies of security on weaker constraints. Second,
because side channels are a serious issue in practical quantum
cryptography. Experimentalists have to be careful that, when they
encode (say) polarization, they encode {\em only} polarization,
and that the device does not change the spectral line, or the
spatial mode, or the temporal mode of the photon as well. Third,
because it is important for practical reasons: quantum
cryptography is becoming a commercial product. If a security
expert recommends a quantum cryptography device, he should be able
to assess that the device acts as it should with "reasonable"
means. After all, the eavesdropper Eve could be herself the
provider of the device!

Anyone faced with this scenario feels at first that, if Eve is
allowed to sell you the devices and you cannot know them in
detail, there is no hope for security. Surprisingly, recent
advances in quantum information suggest that this despair,
reasonable as it is, may be too pessimistic. Let's see where the
hope lies, and which assumptions are really crucial.

\begin{figure}
\includegraphics[width=8cm]{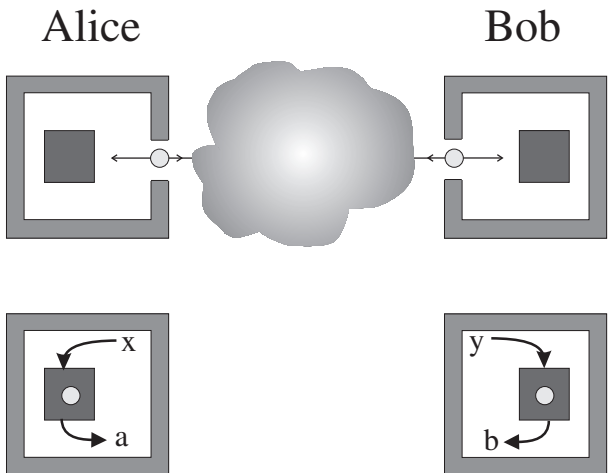}
\caption{A pictorial description of the no-signalling assumption
in our context. The dark grey boxes in Alice's and Bob's
laboratories are the devices provided by Eve. In a first step, the
laboratories are open for the signal that correlates them (grey
spheres). The arrows on the channel indicate that it is not
important whether this signal comes from outside, or is emitted by
Alice's device to Bob's, or viceversa: in any case, it is under
Eve's control. What is important, is that the inputs ($x$, $y$)
have not been chosen yet. In a second step, the laboratories are
absolutely closed: no leakage of information about the inputs
($x$, $y$) or the outputs ($a$, $b$) is allowed. In a third step
(not shown), Alice and Bob can carry out the usual procedures of
error correction and privacy amplification by communicating on an
authenticated channel.}\label{figlabos}
\end{figure}

The scheme to distribute correlations we have in mind is
represented in Fig.~\ref{figlabos}. In Alice's and Bob's
laboratories, the dark grey square represents the device possibly
provided by Eve. The distribution of correlations is made in three
steps. In the {\em first step}, both laboratories are open to the
signal that correlate them. This signal comes either from outside,
or is emitted by Alice's device to Bob's, or viceversa: in any
case, it must be assumed to be under Eve's full control. In the
{\em second step}, the laboratories are completely sealed, an
obviously necessary condition as we are going to see. On the
device that reads the signal, Alice and Bob must have a knob,
which allows them to choose among at least two alternatives (in
usual QKD, this is for instance the choice of the basis). It is
obviously necessary to assume that no information about the
position of the knob leaks out of Alice's and Bob's laboratories
(in QKD, if Eve would know the basis, she can measure the state
without introducing errors). Now, conditioned on the choice of an
input (a position of the knob, labelled $x$ for Alice and $y$ for
Bob), an output is produced ($a$ for Alice, $b$ for Bob). The
lists of $a$ and $b$ constitute the raw key. How can there be some
secrecy in this raw key? The insight from quantum physics is that
the outputs may be {\em not} under the provider's control: if the
probability distribution of the outputs violates some kind of Bell
inequality, then by definition those outputs have not been
produced by shared randomness --- in other words, the correlations
have been produced by the measurements themselves, and did not
pre-exist to them. They could have been produced by communication,
if information about the inputs $x$ and/or $y$ would have
propagated between Alice and Bob; but we have insisted on the
no-signalling assumption: no information about $x$ and $y$ should
leak out of Alice's and Bob's laboratories respectively
\cite{notec}. The {\em third step} is usual: Alice and Bob can
make classical data processing in order to distill a fully secret
key.

The reasoning above is exactly the intuition that led Ekert to
discover (independently of previous works) quantum cryptography in
1991 \cite{ekert}. Ekert's work contains {\em in nuce} the idea of
a {\em device-independent security proof}: it should be possible
to demonstrate that a probability distribution, which violates
some Bell inequality, is secure by this very fact, without any
reference to the formalism of quantum physics. Of course, in
physics as we know it today, a Bell inequality can only be
violated with entanglement: that is why people immediately used
the quantum formalism to study Ekert's intuition \cite{mayers}.
But recently, tools have been developed, that allow one to study
no-signalling distributions in themselves, without the formalism
of Hilbert spaces. It is then possible to come back to the
original intuition by Ekert, and try and prove security only
through the violation of a Bell-type inequality. This is the theme
of the present paper.

Since we need to introduce in more detail the tools used in this
work, we do this in Section \ref{secgen} and postpone the outline
of the paper to paragraph \ref{ssecout}.

\section{Cryptography in the no-signalling polytope}
\label{secgen}

This first section introduces the language and the tools which are
needed in a general framework. We focus from the very beginning
onto {\em bipartite correlations}, i.e. correlations involving two
partners, traditionally called Alice and Bob.

\subsection{Formalization of Bell-type experiments}

The physical situation one must keep in mind is a Bell-type
experiment. Alice and Bob receive several pairs of entangled
quantum particles. On each particle, Alice performs the
measurement $x$ randomly drawn from a finite set of $m_A$
possibilities; as a result, she obtains the output $a$ out of a
discrete set containing $n_A$ symbols. Independently from Alice,
Bob performs the measurement $y$ randomly drawn from a finite set
of $m_B$ possibilities; as a result, he obtains the output $b$ out
of a discrete set containing $n_B$ symbols. Such an experiment is
characterized by the family of probabilities \ba
P(a,b,x,y)&=&P(a,b|x,y)\,P(x)\,P(y)\,. \ea There are
$D\,=\,m_A\,m_B\,n_A\,n_B$ such numbers, so each experiment can be
described by a point in a $D$-dimensional space; more precisely,
in a region of such a space, bounded by the conditions that
probabilities must be positive and sum up to one. By imposing
further restrictions on the possible probability distributions,
the region of possible experiment shrinks, thus adding non-trivial
boundaries \cite{tsi,pito,barrett}. For our study, three
restrictions are meaningful.

The first restriction is the requirement that the probability
distribution must be built without communication, only with shared
randomness. In the literature, this has been known as the
hypothesis of local hidden variables. In our context, these
variables are not hidden "in nature" (as in the original
interpretational debates about quantum physics): they may rather
be hidden in Alice's and Bob's laboratories, in the devices that
Eve has provided to them. The bounded region, which contains all
probability distributions that can be obtained by shared
randomness, forms a polytope, that is a convex set bounded by a
finite number of hyperplanes ("facets"); therefore we refer to it
as to the {\em local polytope}. The vertices of the local polytope
are the points corresponding to {\em deterministic strategies},
that is, strategies in which $a=a(x)$ and $b=b(y)$ with
probability one; that is,
$P(a,b|x,y)=\delta_{a,a(x)}\delta_{b,b(y)}$. There are clearly
$m_A^{n_A}\,m_B^{n_B}$ such strategies. The vertices are thus
easily listed, but to find the facets given the vertices is a
computationally hard task. The importance of finding the facets is
pretty clear. If a point, representing an experiment, lies within
the polytope, then there exists a strategy with shared randomness
(a local variable model) that produces the same probability
distribution. If on the contrary a point lies outside the local
polytope, then the experiment cannot be reproduced with shared
randomness only. The interpretation of the facets of the local
polytope is therefore obvious: they correspond to Bell's
inequalities. We shall call {\em non-local region} the region
which lies outside the local polytope.

The second restriction is the requirement that the probability
distribution must be obtained from measurements on quantum
bipartite systems. The bounded region thus obtained shall be
called the {\em quantum region}. It is not a polytope, since there
is not a finite set of extremal points. It is a convex set if one
really allows all possible measurements on all possible states in
arbitrary-dimensional Hilbert space \cite{pito,ww}; if one
restricts to the measurements on a given state, or even to
von-Neumann measurements on a Hilbert space with given dimension,
convexity is not proved in general (although no counter-example is
known, to our knowledge). Needless to recall, the quantum region
contains the local polytope, but is larger than it: measurement on
quantum states can give rise to non-local correlations (Bell
inequalities are violated).

The third restriction is the requirement that the probability
distribution must not allow signalling from Alice to Bob or
viceversa. The no-signalling requirement is fulfilled if and only
if Alice's marginal distribution does not depend on Bob's choice
of input, and viceversa: that is, the probability distributions
must fulfill
\ba \sum_{b}P(a,b|x,y)&=&P(a|x)\,,\\
\sum_{a}P(a,b|x,y)&=&P(b|y)\,. \ea These conditions define again a
polytope, the {\em no-signalling polytope}, which contains the
quantum region. The deterministic strategies are still vertices
for this polytope; to these, one must add other vertices which
represent, loosely speaking, purely non-local no-signalling
strategies. These additional points, sometimes called non-local
machines or non-local boxes, have been fully characterized only in
a few cases.

\subsection{Secrecy of probability distributions}

Here is the question that we are going to address in this paper.
Alice and Bob have repeated many times the "measurement" procedure
and share an arbitrary large number of realizations of the random
variables distributed according to $P(a,b|x,y)$. By revealing a
fraction of their lists, they can estimate whether their
probability distribution lies in the local polytope or in the
non-local region. The goal is to study whether Alice and Bob can
extract secrecy out of their data with this knowledge only.

To motivate the question, let us consider the best-known quantum
cryptography protocol, the one invented by Bennett and Brassard in
1984 (BB84) \cite{bb84}. In this protocol, $a,b\in\{0,1\}$ and
$x,y\in\{X,Z\}$ are both binary. In the absence of any error, the
BB84 protocol distributes perfect correlations when $x=y$ and no
correlations when $x\neq y$, that is:
$P(0,0|X,X)=P(1,1|X,X)=\demi$, $P(0,0|Z,Z)=P(1,1|Z,Z)=\demi$, and
$P(a,b|X,Z)=p(a,b|Z,X)=\frac{1}{4}$. If Alice and Bob have
obtained their results by measuring two-dimensional quantum
systems (qubits), such correlations provide secrecy under the
usual assumption that the eavesdropper is limited only by the laws
of quantum physics \cite{qbb84}. However, this distribution can
also be obtained with shared randomness: if Alice and Bob would
share randomly distributed pairs of classical bits $(r_X,r_Z)$,
they simply have to output $r_Z$ (respectively $r_X$) if they are
asked to measure $Z$ (respectively $X$). Thus we see the
importance of the additional assumption on the physical
realization, namely, that both Alice and Bob are measuring a
qubit, and therefore the pair $(r_X,r_Z)$ is not available because
$[X,Z]\neq 0$.

In other words, the correlations of BB84, even in the absence of
errors, are not secure "by themselves": they are secure only
provided the quantum degrees of freedom are under good control.
The question we raised can now be put in its true perspective: are
there correlations that are secure by themselves, by the very fact
of being what they are, without having even to describe how Alice
and Bob managed to obtain them from a real channel?

It turns out that it is easier to tackle this question by
considering that {\em the eavesdropper Eve is not even limited by
quantum physics, but only by the no-signalling constraint}. This
means that Eve can distribute any many-instances probability
distribution $P(\vec{a},\vec{b}|\vec{x},\vec{y})$ that lies within
the no-signalling polytope; Alice and Bob have the freedom of
choosing their sequence of measurements ($\vec{x}$ and $\vec{y}$
respectively) and will obtain the corresponding outcomes. By
making this assumption, we stand clearly on the conservative side:
if we can demonstrate that a non-vanishing secret key can be
extracted against such a powerful eavesdropper, then the secret
key achievable against a "realistic" (i.e., quantum) eavesdropper
will be at least as long.

In quantum cryptography, secrecy relies on entanglement. On which
physical quantity can such a strong security, as the one we are
asking for, rely? The answer is: on {\em the non-locality of the
correlations}, that is, on the fact that the correlations cannot
be obtained by shared randomness \cite{notec}. No secrecy can be
extracted if Alice and Bob share a probability distribution which
lies within the local polytope, just as no secrecy against a
quantum Eve can be extracted out of separable states
\cite{lew,ag}.

\subsection{Individual eavesdropping strategies}
\label{individ}

Barrett, Hardy and Kent \cite{bhk} have shown an example of a
protocol, in which quantum correlations can provide secrecy
against the most powerful attack by a no-signalling Eve. This is
the first example that one can achieve security even against a
supra-quantum Eve, showing that security in key distribution
arises from general features of no-signalling distributions rather
than from the specificities of the Hilbert space structure.
However, their example has important limitations: actually, it
provides a protocol to distribute a {\em single} secret bit
(thence zero key rate) in the case when Alice and Bob share
correlations that can be ascribed to {\em noiseless} quantum
states.

In this paper, we tackle the problem from the other side: we don't
go straight for security against the most powerful adversary, but
we follow the same path that was followed historically by quantum
cryptography, namely, {\em we limit the eavesdropper to adopt an
individual strategy}. This means the following: Eve follows the
same procedure for each instance of measurement --- that is, she
is not allowed to correlate different instances. Moreover, Eve is
asked to put her input $z$ before any error correction and privacy
amplification. Consequently, any individual attack is described of
a three-partite probability distribution $P(a,b,e|x,y,z)$ such
that \ba P(a,b|x,y)&=&\sum_e P(e|z)\,P(a,b|x,y,e,z)\,.
\label{eve}\ea Note that Eve is also limited by no-signalling,
that is why the left hand side does not depend on $z$. One can see
that this is an individual attack by looking at it as follows:
when Eve gets outcome $e$ out of her input $z$, she sends out the
point $P(a,b|x,y,e,z)$.

Now we demonstrate two similar, important results about individual
eavesdropping strategies:

{\bf Theorem 1:} Eve can limit herself in sending out {\em
extremal points} of the no-signalling polytope.

{\em Proof.} Suppose that an attack is defined, in which one of
the $P(a,b|x,y,e,z)$ is not an extremal point. Then, this point
can be itself decomposed on extremal points: $P(a,b|x,y,e,z)=
\sum_{\lambda}P(\lambda)P(a,b|x,y,e,z,\lambda)$ where the
$P(a,b|x,y,e,z,\lambda)$ are all extremal. But the knowledge of
$\lambda$ must be given to Eve: by redefining Eve's symbol as
$(e,\lambda)\rightarrow e$, we have an attack which is as powerful
as the one we started from, and is of the form (\ref{eve}) while
having only extreme points in the decomposition.

{\bf Theorem 2:} Suppose that Alice and Bob can transform
$P(a,b|x,y)$ into $\tilde{P}(a,b|x,y)$ by using only local
operations and public communication independent of $a,b,x,y$. Then
there exist a purification of $\tilde{P}(a,b|x,y)$ that gives Eve
as much information as the best purification of $P(a,b|x,y)$.

{\em Proof.} Suppose (\ref{eve}) is the best purification of $P$
from Eve's point of view; for clarity, let's use Theorem 1 to say
that the $P(a,b|x,y,e,z)$ are extremal points. The procedure of
Alice and Bob can be described as follows: for each realization of
the variables $(a,b,x,y)$, Alice draws a random number and reveals
publicly its value $j$; then, she and Bob apply the local
transformation $T_j$ on which they have previously agreed,
transforming $x\rightarrow X_j$, $a\rightarrow A_j$ etc. Since
there is no correlation between $j$ and $(e,z)$, each extremal
point $P_{\epsilon}(a,b|x,y,e,z)$ is transformed into \ba
\tilde{P}(a,b|x,y,e,z) &=&\sum_j P(j)P(A_j,B_j|X_j,Y_j,e,z)\nonumber\\
&=& \sum_{j,\epsilon} P(j)P(\epsilon|e,j)P(a,b|x,y,\epsilon,z)\,.
\ea Consequently, $\tilde{P}(a,b|x,y)$ is a mixture of the
extremal points $P(a,b|x,y,\epsilon,z)$ with weight
$P(\epsilon|z)=\sum_{j,e}P(e|z)P(j)P(\epsilon|e,j)$. To conclude
the proof, just notice that Eve has been able to follow the full
procedure, because she has learnt $j$ and the list of the $T_j$ is
publicly known. Thus, there exist a decomposition of
$\tilde{P}(a,b|x,y)$ onto extremal points that gives Eve as much
information as the best decomposition of ${P}(a,b|x,y)$.

\subsection{Outline of the paper}
\label{ssecout}

This is all that could be said in full generality. In what
follows, we study mainly scenarios in which $x,y\in\{0,1\}$: Alice
and Bob choose between two possible measurements. In Section
\ref{sec2}, we address the case where also the outcomes $a$ and
$b$ are both binary; apart from paragraph \ref{sec2unc}, all the
results of this Section have been announced in Ref.~\cite{prl}. In
Section \ref{secd}, we explore the case where both the outcomes
are $d$-valued, in particular for $d=3$. In both situations, we
shall consider an explicit protocol for Alice and Bob, without
claim of optimality. Conclusions and perspectives are listed in
Section \ref{secpersp}.

\section{Binary Outcomes}
\label{sec2}

In this Section, we consider $m_A=m_B=n_A=n_B=2$; that is,
$a,b,x,y\in\{0,1\}$ are all binary. Below, all the sums involving
bits are to be computed modulo 2.

\subsection{The polytopes and the quantum region}

In the case of binary inputs and outputs, the local and the
no-signalling polytopes have been fully characterized, and their
structure is rather simple. A lot (but not all) is known about the
quantum region too. Under no-signalling, the full probability
distribution is entirely characterized by eight probabilities,
therefore all these objects live in an 8-dimensional space.

The {\em local polytope} \cite{fine,collins} has eight non-trivial
facets. Up to symmetries like relabelling of the inputs and of the
outputs, they are all equivalent to the Clauser-Horne-Shimony-Holt
(CHSH) inequality \cite{chsh}. The representative of this
inequality reads \ba CHSH&=&\sum_{x,y=0}^1 P(a+b=xy|xy)\leq 3\,.
\label{chsh}\ea On each facet lie eight out of the sixteen
deterministic strategies; these are said to {\em saturate} the
inequality, because by definition they give $CHSH=3$. Note that
the eight points on a facet are linearly independent from one
another \cite{note0}. The deterministic strategies that saturate
our representative (\ref{chsh}) are readily seen to be the
following ones: \ba
\begin{array}{lcl} L_1^r&=&
\{a(x)=r,b(y)=r\}\\L_2^r&=&\{a(x)=x+r,b(y)=r\} \\ L_3^r&=&
\{a(x)=r,b(y)=y+r\}\\L_4^r&=&\{a(x)=x+r,b(y)=y+r+1\}
\end{array}\label{eightlocs}\ea where $r=0,1$.

The {\em no-signalling polytope} \cite{barrett} is obtained from
the local polytope by adding a single extremal non-local point on
top of each CHSH facet. The non-local point on top of our
representative is defined by \ba PR &=&
\demi\,\delta(a+b=xy)\,.\label{prbox}\ea This point is the
so-called {\em PR-box}, invented by Popescu and Rohrlich \cite{pr}
and by Tsirelson \cite{tsi}. It violates the CHSH inequality up to
its algebraic limit $CHSH=4$.

About the {\em quantum region}: the set of {\em correlations} that
are producible by measuring quantum states is known, and
corresponds to what can be produced by measurement on two-qubit
states \cite{cirelson80}. It is an open question, whether the
analysis of the marginals can reveal further features of the
quantum region. The violation of CHSH is bounded by \ba
CHSH(QM)&\leq & 2+\sqrt{2}\,, \ea  where the maximum is reached
with the probability distribution \ba
P(a+b=xy|xy)&=&\frac{1+\frac{1}{\sqrt{2}}}{2} \label{probaq}\ea
obtained by measuring suitable observables on a maximally
entangled state.

\subsection{The non-local raw probability distribution}

To study the possibility of secret key extraction, we can restrict
our attention to the sector of the non-local region that lies
above a given facet of the local polytope, say the representative
one for which we have collected the tools above. Any point in this
sector, by definition, can be decomposed as a convex combination
of the PR-box (\ref{prbox}) and of the eight deterministic
strategies on the facet (\ref{eightlocs}). As shown above, we can
assume without loss of generality that Eve distributes these nine
strategies. We shall write $p_{NL}$ (for "non-local") the
probability that Eve sends the PR-box to Alice and Bob; and
$p_j^r$ the probability that Eve sends the deterministic strategy
$L_j^r$. We shall also write $p_L=\sum_{j,r}p_j^r=1-p_{NL}$.

The statistics generated by Eve sending the extremal points are
summarized in the Table \ref{tableraw}. The reading of this Table
is pretty clear. For instance, one finds that
$P(a=b=0|x=y=0)=p_{NL}/2+p_1^0+p_2^0+p_3^0$. To obtain the
$P(a,b,x,y)$, one must multiply the entries of the Table by
$P(x)P(y)$. Since we are supposing that the extremal points are
sent to Alice and Bob by Eve, the label of each point can be
considered also as Eve's symbol.

Finally, we note that using Table \ref{tableraw} in (\ref{chsh}),
one finds \ba p_{NL}&=& CHSH\,-\,3\,. \ea In other words, $p_{NL}$
measures directly the violation of the CHSH inequality. It follows
that in the quantum region \ba p_{NL}(QM)&\leq &
\sqrt{2}-1\,\approx 0.414\,. \label{pnlq}\ea

\subsection{The CHSH protocol for cryptography}
\label{chshproto}

Whenever Eve distributes the PR-box, she has no information at all
about the bits received by Alice and Bob, because of the monogamy
of those correlations \cite{barrett}. On the contrary, when she
distributes a deterministic strategy, she has some information,
depending on the actual cryptographic protocol. The question is
thus, which is the best procedure to extract a secret key out of
the raw distribution of Table \ref{tableraw}? We have no answer in
full generality; but we can notice a few things and propose a
protocol which is a reasonable candidate for optimality. A good
cryptography protocol should (i) present high correlations between
Alice and Bob, and (ii) reduce Eve's information as much as
possible. Now, in the raw data, we see that Alice and Bob are
highly anti-correlated when $x=y=1$: it is thus natural to devise
a procedure that allows them to transform these anti-correlations
in correlations. A good procedure reveals as small as possible
information on the public channel.

The protocol we propose, and that we call {\em CHSH protocol} for
obvious reasons, is the following:
\begin{enumerate}
\item {\em Distribution.} Alice and Bob repeat the measurement
procedure on arbitrarily many instances and collect their data.

\item {\em Parameter estimation.} By revealing publicly some of
their results, they estimate the parameters of their distribution,
in particular the fraction $p_{NL}$ of intrinsically non-local
correlations.

\item {\em Pseudo-Sifting.} For each instance, Alice reveals the
measurement she has performed ($x=0$ or $x=1$). Whenever Alice
declares $x=1$ and Bob has chosen $y=1$, Bob flips his bit. Bob
does not reveal the measurement he has performed. This is the
procedure which transforms anti-correlations into correlations
while revealing the smallest amount of information on the public
channel. We call it pseudo-sifting, because it enters in the
protocol at the same place as sifting occurs in other protocols,
but here all the items are kept.

\item {\em Classical processing.} The details depend on whether
one considers one-way post-processing ("error correction and
privacy amplification", efficient in terms of secret key rate) or
two-way post-processing ("advantage distillation", inefficient for
small errors but tolerating larger errors). The two cases are
discussed separately below.

\end{enumerate}

After pseudo-sifting, and writing $\xi_{j}=P(y=j)$, we can write
the Alice-Bob-Eve distribution splits into two, one for each value
of $x$, as given in Table \ref{tablesx}. It is important to
understand the content of these two distributions. Suppose Eve has
sent out $L_1^0$ and Alice has announced $x=0$ in the
pseudo-sifting phase. Then Eve knows for sure both Alice's and
Bob's outcomes, here $a=0$ and $b=0$; and in fact, for $x=0$, the
strategy $L_1^0$ gives only this result. However, if Eve has sent
out $L_3^0$ and Alice has announced $x=0$, things are different:
Eve still knows for sure Alice's outcome ($a=0$), but Bob's
outcome depends on his input, being $b=0$ if $y=0$, $b=1$ if
$y=1$. Remarkably, the roles are exactly reversed for $x=1$: in
this case, $L_3^0$ produces only $a=0$ and $b=0$; while $L_1^0$
gives $a=0$, $b=0$ if $y=0$ and $b=1$ if $y=1$.

In fact, a closer examination of the Tables shows that all the
eight local point have such a behavior: (i) Alice's outcome $a$ is
always known to Eve, because the setting used by Alice is publicly
known. (ii) If a local point provides Eve with full information
about Bob's outcome $b$ when $x=0$, the same point leaves her
uncertain about $b$ when $x=1$; and viceversa. We shall come back
to this interesting feature in paragraph \ref{sec2unc}. Obviously,
Eve's uncertainty is maximal when Alice's and Bob's settings are
chosen at random, therefore we set from now on \ba
P(x=i)\,=\,P(y=j)\,\equiv\,\,\xi_j&=&\demi\,. \label{pdemi}\ea
Note that this situation is different from quantum cryptography:
in the quantum case, Eve's information does not depend on the
frequency with which each setting is used, and in fact Alice and
Bob can use almost always the same setting, provided they use the
other one(s) sometimes in order to check coherence
\cite{qkdasymm}.

Now we can understand better the advantage of our pseudo-sifting
procedure. If neither Alice nor Bob would reveal their setting,
Eve's information on the deterministic strategies would decrease,
but Alice and Bob would stay anti-correlated when $x=y=1$. If on
the contrary both Alice and Bob would reveal their setting, Eve
would have full information on both $a$ and $b$ for every
deterministic strategy. The pseudo-sifting procedure corrects for
the anti-correlation, and keeps some uncertainty in Eve's
knowledge about Bob's result.

In summary, Table \ref{tablesx} contains the probability
distribution Alice-Bob-Eve after pseudo-sifting. Now we must study
whether one can extract secrecy out of them, using classical pre-
and post-processing. Before turning our attention to that, we want
to stress a nice feature of the distribution we have just
obtained.

\subsection{Uncertainty relations}
\label{sec2unc}

Remarkably, the protocol we have defined exhibits a feature which
is also present in quantum cryptography, namely the fact that Eve
gains information on a "basis" at the expense of introducing
errors in the complementary one. Here it is precisely.

Refer to Table \ref{tablesx}, recalling that $\xi_0=\xi=\demi$.
The probabilities $p(a\neq b|x)$ of error between Alice and Bob
when $x=0$ and $x=1$ are respectively
\ba e_{AB|0}&=&\demi\,\left(p_3^0+p_3^1+p_4^0+p_4^1\right)\,,\\
 e_{AB|1}&=&\demi\,\left(p_1^0+p_1^1+p_2^0+p_2^1\right)\,. \ea
Eve's uncertainty on Bob's symbol, measured by conditional Shannon
entropy, is \ba
H(B|E,x=0)&=& 1-\left(p_1^0+p_1^1+p_2^0+p_2^1\right)\\
H(B|E,x=1)&=& 1-\left(p_3^0+p_3^1+p_4^0+p_4^1\right)\,. \ea Thus,
there appear in our protocol a cryptographic {\em uncertainty
relation} in the form \ba H(B|E,x)&=&1-2e_{AB|x+1}\,. \ea The
origin of this relation is rather clear. The pseudo-sifting phase
of the protocol is optimized to extract correlations from the
non-local strategy (PR-box), but on deterministic strategies, the
pseudo-sifting has another action. Specifically: for $L_{1}^r$ and
$L_2^r$, after pseudo-sifting we have $b(y=0)=b(y=1)=a$ when $x=0$
(no error, and Eve knows $b$), and $b(y=0)\neq b(y=1)$ when $x=1$
(error in half cases, and Eve does not know $b$); for $L_{3}^r$
and $L_4^r$, it's just the opposite. In summary, for each local
strategy, Eve learns everything only for one Alice's setting, and
for the other an error between Alice and Bob occurs half of the
times.

This is the first evidence of an analogue of quantum mechanical
uncertainty relations in a generic no-signalling theory. We can
now move to the main issue, the extraction of a secret key.

\subsection{One-way classical post-processing}
\label{sec2r1w}

\subsubsection{Generalities}

For one-way classical post-processing, the bound for the length of
the achievable secret key rate under the assumption of individual
attacks is the Csisz\'ar-K\"orner (CK) bound \cite{CK,AC}. In the
case where Eve's knows more about Alice's symbol than about Bob's,
as is the case here, the CK bound reads \ba
R_{CK}&=&\sup_{(B',T)\leftarrow B}\,\big[ H(B'|E,T)-H(B'|A,T)\big]
\ea where $B\rightarrow (B',T)$ is called {\em pre-processing}:
from his initial data $B$, Bob obtains some processed data $B'$
that he does not reveal, and some other processed data $T$ that
are broadcasted on a public channel. For classical distributions,
bitwise pre-processing is already optimal \cite{CK,AC}. In this
paper, we have not explored the possible use of $T$: in this case,
the pre-processing reduces to flipping each bit with some
probability $q$. Consequently, we'll have an estimate $r_{CK}\leq
R_{CK}$ for the achievable secret key rate. Recalling the link
$I(X:Y)=H(X)-H(X|Y)$ between Shannon entropies and mutual
information, we write our estimate for the CK bound as \ba
r_{CK}&=&\max_{B'\leftarrow B}\,\big[
I(A:B')-I(B':E)\big]\nonumber\\&=&\demi\,\sum_{x=0,1}
\max_{B'\leftarrow B}\,\big[ I(A:B'|x)-I(B':E|x)\big]\,. \ea

Let's sketch the computation explicitly for $x=0$ (for
conciseness, we omit to write this condition in the formulae
below). In Table \ref{tablesx}, one reads for $p(a,b)$: \ba
\begin{array}{lcl}
p(0,0)&=&\frac{p_{NL}}{2}+p_1^0+p_2^0+\frac{p_3^0+p_4^0}{2}\\
p(0,1)&=&\frac{p_3^0+p_4^0}{2}\\
p(1,0)&=&\frac{p_3^1+p_4^1}{2}\\
p(1,1)&=&\frac{p_{NL}}{2}+p_1^1+p_2^1+\frac{p_3^1+p_4^1}{2}
\end{array}\,.
\ea If we denote by $q$ the probability that Bob flips his bit in
the pre-processing, then \ba p(a,b')&=&(1-q)p(a,b=b')+q
p(a,b=b'+1)\,. \ea These four probabilities allow to compute the
mutual information $I(A:B')=H(A)-H(A|B')$. Turning to Eve: before
pre-processing, she has full knowledge on Bob's symbol for $L_1^r$
and $L_2^r$, and no knowledge for $L_3^r$ and $L_4^r$. As a
consequence of the fact that Eve knows exactly on which items she
has full information and on which she has no information at all,
one has simply \ba
H(B'|E)&=&H(B|E)+\,\big[1-H(B|E)\big]h(q)\label{ie0} \ea where $h$
is binary entropy. The calculation is of course identical for
$x=1$ and this allows to compute $r_{CK}$ for any probability
distribution. We focus explicitly on two cases.

\subsubsection{Isotropic distribution}
\label{ssiso}

Let's consider an isotropic probability distribution, that is, a
distribution of the form \ba
P(a,b|x,y)&=&\frac{1+p_{NL}}{4}\,\delta(a+b=xy)+\frac{p_L}{8}\,.
\ea This necessarily implies $p_j^r=p_L/8$ for all $j,r$, since
recall that the $L_j^r$ are linearly independent. Note that the
point of highest violation in the quantum region (\ref{probaq}) is
of this form, with $p_{NL}=\sqrt{2}-1$.

Remarkably, Alice and Bob can transform any distribution with a
given $p_{NL}$ to the isotropic distribution defined by the same
$p_{NL}$ with local operations and public communication, a
procedure called "depolarization" \cite{generic}. This implies
that the results of this paragraph are in some sense generic. In
fact, by Theorem 2 of paragraph \ref{individ}, Eve's best
individual eavesdropping strategy for a fixed value of $p_{NL}$
consists in preparing an isotropic distribution. Alternatively, we
can modify the protocol to add the fact that Alice and Bob apply
systematically the depolarization procedure.

For isotropic distributions, the two tables for $x=0$ and $x=1$
become identical, and we can rewrite them as Table \ref{tableiso}.
In this Table, we have changed the notation for Eve's knowledge,
and have written $(a,b)$ when Eve knows both outcomes, $(a,?)$
when she knows only Alice's, and $(?,?)$ when she knows none.

This distribution has $p(a=0)=p(a=1)=\demi$. Before
pre-processing, the error between Alice and Bob is $e_{AB}=p_L/4$;
after pre-processing, the quantity to be corrected in error
correction is $e'_{AB}=(1-q)e_{AB}+q(1-e_{AB})$. Eve's information
is $\frac{p_L}{2}[1-h(q)]$. Thus \ba
r_{CK}&=&\max_{q\in[0,\demi]}\,\left[1-h(e'_{AB})-\frac{p_L}{2}[1-h(q)]\right]\,.
\ea This quantity is plotted in Fig.~\ref{figqkd} as a function of
the disturbance $D$ defined by $p_{NL}=\sqrt{2}(1-2D)-1$. This
parameter characterizes the properties of the channel linking
Alice and Bob: it is therefore useful for comparison with a
quantum realization of the CHSH protocol and with BB84, see
\ref{sec2rq} and Appendix \ref{appcrypto}. We see that $r_{CK}>0$
for $D\lesssim 6.3\%$ that is $p_{NL}\gtrsim 0.236$ for the
optimal pre-processing. Without pre-processing, the bound becomes
$p_{NL}\gtrsim 0.318$. The important remark is that both these
values are {\em within the quantum region} (\ref{pnlq}). This
means that using quantum physics, one can distribute correlations
which allow (at least against individual attacks) the extraction
of a secret key without any further assumption about the details
of the physical realization.

\subsubsection{Reaching the Bell limit}

Another interesting example deals with the following question: can
one find one-parameter families of probability distributions for
which $R_{CK}>0$ as soon as $p_{NL}>0$; that is, distributions for
which one can extract a secret key out of one-way processing, down
to the limit of the local polytope? The answer is yes, and this
can be achieved even without pre-processing. Here is an example:
set $p_1^0=p_2^0$, $p_1^1=p_2^1$, and $p_{3,4}^r=0$. For both
$x=0$ and $x=1$ we have $p(a=0)=p(a=1)=\demi$. For $x=0$, Alice
and Bob make no errors ($e_{AB|0}=0$), and Eve's information is
$I(B:E|x=0)=p_L$; for $x=1$, the errors of Alice and Bob are
$e_{AB|1}=\frac{p_L}{2}$ and Eve has no information. In summary,
even neglecting pre-processing, \ba
r_{CK}&=&1-\demi\,h(p_L/2)-\frac{p_L}{2} \ea which is strictly
positive in the whole region $p_L<1$.

Note that the distributions described here cannot be broadcasted
using quantum states. The reason is that the quantum intersection
with the non-local region is strictly inside this region, where
"inside" means that, as soon as $p_{NL}>0$, all the $p_j^r$ must
be non zero, because the $L_j^r$ are linearly independent. On the
contrary, here we have set $p_{3,4}^r=0$. Anyway, in spite of the
fact that we are not able to broadcast this distribution with
known physical means, it is interesting to notice that there
exists a family of probability distributions that can lead to a
secret key under one-way post-processing, for any amount of
non-locality.

\subsection{Two-way classical post-processing}
\label{sec2r2w}

\subsubsection{Advantage distillation (AD)}

Contrary to the one-way case, no tight bound like the
Csisz\'ar-K\"orner bound is known when two-way classical
post-processing is allowed; nor is the optimal procedure known.
The best-known two-way post-processing is the so-called {\em
advantage distillation (AD)}. Forgetting about pre-processing, one
can see the effect of AD as follows: starting from a situation
where $I(A:B)<I(B:E)$, one makes a processing at the end of which
the new variables satisfy
$I(\tilde{A}:\tilde{B})>I(\tilde{B}:\tilde{E})$; at this point,
one applies the one-way post-processing.

In AD, Alice reveals $N$ instances such that her $N$ bits are
equal: $a_{i_1}=...=a_{i_N}=\alpha$. Bob looks at the same
instances, and announces whether his bits are also all equal. If
indeed $b_{i_1}=...=b_{i_N}=\beta$, which happens with probability
$(1-e_{AB})^N+e_{AB}^N$, Alice and Bob keep one instance;
otherwise, they discard all the $N$ bits. Bob's error on Alice's
symbols becomes \ba
\tilde{e}_{AB}&=&\frac{e_{AB}^N}{(1-e_{AB})^N+e_{AB}^N}\,\approx\,
\left(\frac{e_{AB}}{1-e_{AB}}\right)^N\,. \ea Notice that
$\tilde{e}_{AB}\rightarrow 0$ in the limit $N\rightarrow\infty$:
this means that $\alpha=\beta$ almost always, for $N$ sufficiently
large. This remark is used to estimate Eve's probability of error
(see below for concrete applications). Typically, one finds that
Eve's error on Bob's symbols goes as \ba \tilde{e}_{E}&\gtrsim&
C\,\left(f(e_{AB})\right)^N \ea with $f(.)$ some function which
depends on the probability distribution under study. Now, as long
as the condition \ba f(e_{AB})&>& \frac{e_{AB}}{1-e_{AB}}
\label{critad}\ea is fulfilled, Eve's error at the end of AD is
exponentially larger than Bob's for increasing $N$: there exists
always a finite value of $N$ such that Eve's error becomes larger
than Bob's. The bound on the tolerable error after AD is then
computed by solving eq.~\ref{critad}.

We apply this procedure to the isotropic correlations described
above (\ref{ssiso}), first without pre-processing, then by
allowing Alice and Bob to perform some bit flip before starting
AD. We anticipate the result: we find that a key can be extracted
for $p_{NL}\gtrsim 0.09$; that is, even with two way
post-processing we are not able to reach the Bell limit for
isotropic correlations. It is an open question, whether the Bell
limit can be reached by a better two-way post-processing for the
isotropic distribution.

\subsubsection{AD without pre-processing}

We refer to Table \ref{tableiso}. We have, as above,
$e_{AB}=\frac{p_L}{4}$. We must now estimate Eve's error on Bob's
symbol after AD. Eve knows $\alpha$ as soon as she knows one of
Alice's symbols $a_{i_k}$, and recall that asymptotically the
guess $\beta=\alpha$ is correct. The only situation in which Eve
is obliged to make a random guess is therefore the case in which
all the $N$ instances correspond to Eve's symbol $(?,?)$. The
probability that Eve's guess of Bob's symbol is wrong is therefore
\ba \tilde{e}_{E}&\gtrsim&
\demi\,\left(\frac{p_{NL}}{1-e_{AB}}\right)^N \label{bound1}\ea
where the denominator comes from the fact that we must condition
on the bit's acceptance. Using (\ref{critad}), we obtain that
secrecy can be extracted as long as $p_{NL}>p_L/4$ that
$p_{NL}>1/5$. This is lower than the bound obtained for one-way
post-processing, as expected.

\subsubsection{AD with pre-processing}

The previous bound can be further improved by allowing Alice and
Bob to pre-process their lists before starting AD. For two-way
post-processing, it is not known whether bitwise pre-processing is
already optimal; but we restrict to it in this work. Specifically,
we suppose that Alice flips her bit with probability $q_A$, Bob
with probability $q_B$. By inspection, one finds that the
probability distribution obtained from Table \ref{tableiso} after
this pre-processing is the one of Table \ref{tableisopp}, where we
have written $\bar{q}=1-q$. Just by looking at the Table, one can
guess the interest of pre-processing: the five possible symbols
for Eve are now spread in all the four cells of the table. For
instance, Eve's symbol is $(0,0)$ was present only in the case
$a=b=0$ in Table \ref{tableiso}, that is, whenever she had this
symbol Eve had full information; this is no longer the case in
Table \ref{tableisopp}. Note also that the roles of $q_A$ and
$q_B$ are not symmetric, because only $q_A$ mixes the strategies
for which Eve does not know Bob's symbol.

The distribution of Table \ref{tableisopp} is such that \ba
e'_{AB}&=&\left(p_{NL}+\frac{p_L}{2}\right)
\,({q}_A\bar{q}_B+\bar{q}_A{q}_B)\,+\,\frac{p_L}{4}\,. \ea The
estimate of Eve's error requires some attention. As before, we
assume that as soon as Eve guesses correctly Alice's symbol
$\alpha$, she automatically guesses also $\beta$; so the question
is, when is Eve uncertain about $\alpha$, in the asymptotic regime
of large $N$? Of course, inequality (\ref{bound1}) still holds
with $e'_{AB}$ replacing $e_{AB}$; but this condition is too weak
here: it does not make any use of the uncertainty introduced on
Eve's knowledge by the pre-processing.

Eve's situation now is such that, even if she has a symbol $(a,b)$
or $(a,?)$, she cannot be completely sure whether $\alpha=a$ or
not. Suppose that among her $N$ symbols, Eve has $n_0$ times the
symbol $(?,?)$, $n_1^0$ times the symbol $(0,?)$, $n_1^1$ times
the symbol $(1,?)$, $n_2^0$ times the symbol $(0,0)$, and $n_2^1$
times the symbol $(1,1)$. Eve cannot avoid errors when
$p(a=0|e)=p(a=1|e)$, that is when $n_1^0=n_1^1\equiv n_1$ and
$n_2^0=n_2^1\equiv n_2$. We have therefore the bound \ba
\tilde{e}'_E&\gtrsim& \demi\sum_{n_0,n_1,n_2}\frac{N!}{n_0!
(n_1!)^2 (n_2!)^2}\,
\gamma_{(?,?)}^{n_0}\,\gamma_1^{2n_1}\,\gamma_2^{2n_2} \ea where
the sum is taken under the constraint $n_0+2n_1+2n_2=N$,
$\gamma_e$ is the probability that Eve has symbol $e$ conditioned
on the bit's acceptance, and
$\gamma_1\equiv\sqrt{\gamma_{(0,?)}\gamma_{(1,?)}}$,
$\gamma_2\equiv\sqrt{\gamma_{(0,0)}\gamma_{(1,1)}}$. By using
$(n!)^2\sim (2n)!/2^{2n}$ and summing the multinomial expansion,
we obtain \ba \tilde{e}'_E&\gtrsim&
\frac{1}{8}\,\left(\gamma_{(?,?)}
+2\gamma_1+2\gamma_2\right)^N\,.\ea Now we must find the
expressions for the $\gamma_e$ in Table \ref{tableisopp}. Suppose
for definiteness that Alice and Bob have accepted the bit
$\alpha=\beta=0$: this happens with probability
$\frac{1-e'_{AB}}{2}$. The probability that this happens and that
Eve has got the symbol $(?,?)$ is
$\frac{p_{NL}}{2}(\bar{q}_A\bar{q}_B+{q}_A{q}_B)$; whence
$\gamma_{(?,?)}=\frac{p_{NL}(\bar{q}_A\bar{q}_B+{q}_A{q}_B)}{1-e'_{AB}}$.
Similarly, the probability that Alice and Bob accept the bit 0 and
that Eve has got $(0,?)$, respectively $(1,?)$, is
$\frac{p_L}{8}\bar{q}_A$, respectively $\frac{p_L}{8}{q}_A$;
whence $\gamma_{1}=
\frac{p_L\sqrt{\bar{q}_A{q}_A}}{4\,(1-e'_{AB})}$. In a similar
way, one computes $\gamma_2$. By writing
$\delta_e\equiv(1-e_{AB}')\gamma_{e}$, we have then \ba
\begin{array}{lcl} \delta_{(?,?)}&=&
p_{NL}\,(\bar{q}_A\bar{q}_B+{q}_A{q}_B)\,,\\
\delta_{1}&=&\frac{p_L}{4}\,\sqrt{\bar{q}_Aq_A}\,,\\
\delta_{2}&=&\frac{p_L}{2}\,\sqrt{\bar{q}_Aq_A}\,
\sqrt{\bar{q}_Bq_B}\end{array} \ea and the condition for
extraction of a secret key becomes \ba \delta_{(?,?)}+
2\delta_1+2\delta_2&
>& e'_{AB}\,. \ea The optimization over $q_A$
and $q_B$ can be done numerically. The result is that a secret key
can be extracted at least down to $p_{NL}\approx 0.09$.

\subsubsection{Positivity of intrinsic information}
\label{secintr}

Given a tripartite probability distribution, $P(a,b,e)$, an upper
bound to the secret-key rate $R$ is given by the so-called
intrinsic information $I(A:B\downarrow E)$, denoted more briefly
in what follows by $I_\downarrow$. This function, introduced in
\cite{MW}, reads
\begin{equation}
\label{intrinf}
    I(A:B\downarrow E)=\min_{E \rightarrow \bar E}I(A:B|\bar E) ,
\end{equation}
the minimization running over all the channels $E \rightarrow \bar
E$. Here, $I(A:B|E)$ denotes the mutual information between Alice
and Bob conditioned on Eve. That is, for each value of Eve's
variable $e$, the correlations between Alice and Bob are described
by the conditioned probability distribution $P(a,b|e)$. The
conditioned mutual information $I(A:B|E)$ is equal to the mutual
information of these probability distribution averaged over
$P(e)$. The exact computation of the intrinsic information is in
general difficult. However, a huge simplification was obtained in
\cite{CRW}, where it was shown that the minimization in Eq.
(\ref{intrinf}) can be restricted to variables $\bar E$ of the
same size as the original one, $E$. This allows a numerical
approach to this problem.

The intrinsic information can be understood as a witness of secret
correlations in $P(a,b,e)$. Indeed, a probability distribution can
be established by local operations and public communication if,
and only if, its intrinsic information is zero \cite{RW}. It is
then clear why the positivity of the intrinsic information is a
necessary condition for positive secret-key rate. Whether it is
sufficient is at present unknown: strong support has been given to
the existence of probability distributions such that $R=0$ and
$I_\downarrow
>0$. These would constitute examples of probability distributions
containing bound information \cite{GW}, that is non-distillable
secret correlations. The existence of bound information has been
proven in a multipartite scenario consisting of $N>2$ honest
parties and the eavesdropper \cite{multbound}. However, it remains
as an open problem for the more standard bipartite scenario.

Using these tools, it is possible to study the secrecy properties
of the probability distribution $P(a,b,e)$ derived from the
previous CHSH-protocol. A first computation of its conditioned
mutual information gives $I(A:B|E)=p_{NL}$. This result easily
follows from Table \ref{tablesx}: when $e=(?,?)$, that happens
with probability $p_{NL}$, Alice and Bob are perfectly correlated,
so their mutual information is equal to one. In all the remaining
cases, e.g. $e=(0,0)$, Alice and Bob have no correlations. Using
this observation, one can guess the optimal map $E\rightarrow\bar
E$. In order to minimize the conditioned mutual information, this
map should deteriorate the perfect correlations between Alice and
Bob when $e=(?,?)$. A way of doing this is by mapping $(0,?)$ and
$(1,?)$ into $(?,?)$, leaving the other symbols unchanged
\cite{noteintr}. We conjecture that this defines the optimal map
for the computation of the intrinsic information. Actually, all
our numerical evidence supports this conjecture. Thus, the
conjectured value for the intrinsic information is
\begin{equation}\label{iichch}
    I_\downarrow= \left(1-\frac{p_L}{2}\right)
    \left(1-h\left(\frac{p_L}{4-2p_L}\right)\right) .
\end{equation}
Interestingly, this quantity is positive whenever $p_{NL}>0$. If
the conjecture is true, it implies that either (i) it is possible
to have a positive secret-key rate for the whole region of Bell
violation, using a new key-distillation protocol, or (ii) the
probability distribution of Table \ref{tablesx} represents an
example of bipartite bound information for sufficiently small
values of $p_{NL}$.

\subsection{Quantum cryptographic analysis of the CHSH protocol}
\label{sec2rq}

It is interesting to analyze the CHSH protocol with the standard
approach of quantum cryptography: Alice and Bob share a quantum
state of two qubits and have agreed on the physical measurements
corresponding to each value of $x$ and $y$; Eve is constrained to
distribute quantum states, of which she keeps a purification.
Recent advances have provided a systematic recipe to find a lower
bound on the secret key rate, that is, to discuss security when
Eve is allowed to perform the most general strategy compatible
with quantum physics (such bounds have been called "unconditional
security proofs", but it should be clear by all that precedes that
this wording is unfortunate).

\begin{figure*}
\includegraphics[width=8cm]{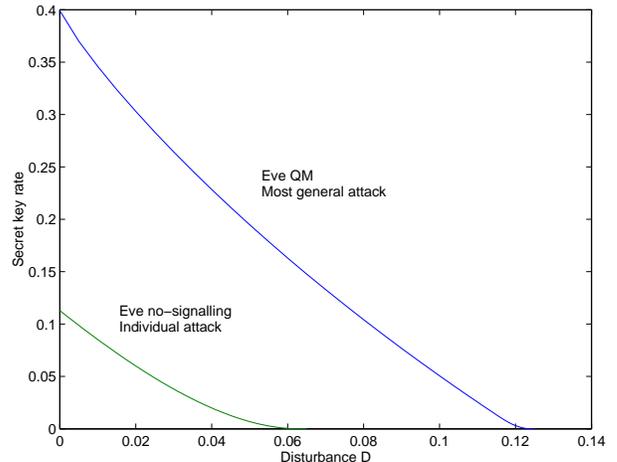}
\caption{Achievable secret key rate for the CHSH protocol, after
one-way post-processing: against a no-signalling Eve for
individual attacks, isotropic distribution (\ref{ssiso}) and
against a quantum Eve, in a two-qubit implementation
(\ref{sec2rq}).}\label{figqkd}
\end{figure*}

The resulting bound on the achievable secret key rate is plotted
in Fig.~\ref{figqkd}. Since the formalism used to compute this
bound is entirely different from the tools used in the present
study, we give this calculation in Appendix \ref{appcrypto}. It
turns out that the CHSH protocol is equivalent to the BB84
protocol plus some classical pre-processing. In particular, the
robustness to noise is the same for both protocols. For low error
rate, BB84 provides higher secret-key rate; however, BB84 cannot
be used for a device-independent proof, since (as we noticed
above) its correlations become intrinsically insecure if the
dimensionality of the Hilbert space is not known.

\section{Larger-dimensional Outcomes}
\label{secd}

In this Section, we explore the generalization of the previous
results to the case of binary inputs and $d$-nary outputs:
$m_A=m_B=2$, $n_A=n_B=d$; that is, $x,y\in\{0,1\}$ and
$a,b\in\{0,1,...,d-1\}$. Below, all the sums involving dits are to
be computed modulo $d$.

For this study, it is useful to introduce a notation for
probability distributions and inequalities \cite{note2}. While the
full probability space is $4d^2$-dimensional, one can verify that
only $4d(d-1)$ parameters are needed to characterize completely a
no-signalling probability distribution --- in other words,
$D=4d(d-1)$ is the dimension of the space in which the
no-signalling and the local polytopes are embedded. We choose the
$\{P(a|x),\,a={0,1,...,d-2};\,x=0,1\}$ ($d-1$ numbers for each
value of $x$), the $\{P(b|y),b={0,1,...,d-2};\,y=0,1\}$ ($d-1$
numbers for each value of $y$), and the $\{P(a,b|x,y),\,a,b=
{0,1,...,d-2};\,x,y=0,1\}$ ($(d-1)^2$ numbers for each value of
$x,y$). This we arrange in arrays as follows:
\ba P&=&\begin{array}{c|c|c|}
A\setminus B & P(b|0)& P(b|1)\\\hline P(a|0)& P(a,b|0,0)&
P(a,b|0,1)\\\hline P(a|1)& P(a,b|1,0)& P(a,b|1,1)\\\hline
\end{array} \ea Note that this array has $2(d-1)$ lines and as
many columns: information on the values $a,b=d-1$ is redundant for
all inputs because of no-signalling. Of course, there is no
problem in working with the "full" array with $2d\times 2d$ if one
finds it more convenient, provided the additional entries are
filled consistently because these parameters are not free.

This notation will also be used for inequalities: in this case,
the numbers in the arrays are the {\em coefficients} which
multiply each probability in the expression of the inequality.
Examples will be provided below.

\subsection{Polytopes and the quantum region}

\subsubsection{Known characterization}

As one might expect, the characterization of the local and the
no-signalling polytopes are an increasingly hard task, as the
dimension of the output increases.

Numerical studies \cite{collins} have provided an unexpectedly
simple structure for the {\em local polytope} for small $d$: as it
happened for $d=2$, all the non-trivial facets appear to be
equivalent to the Collins-Gisin-Linden-Massar-Popescu (CGLMP)
inequality \cite{cglmp,masanes,collins,note3} \ba I_d&=&
\begin{array}{c|cccc|cccc|} A\setminus B& -1&-1&\ldots &-1& 0& 0&\ldots
&0\\\hline -1&  1&1&\ldots &1& 1&0&\ldots &0\\ -1&  0&1&\ldots &1&
1&1&\ldots &0\\ \vdots & \vdots & &\ddots & \vdots& \vdots &
&\ddots & \vdots\\ -1& 0&0&\ldots &1& 1&1&\ldots &1\\\hline 0&
1&0&\ldots &0& -1&0&\ldots &0\\ 0&  1&1&\ldots &0& -1&-1&\ldots
&0\\ \vdots & \vdots & &\ddots & \vdots& \vdots & &\ddots &
\vdots\\ 0& 1&1&\ldots &1& -1&-1&\ldots &-1\\\hline
\end{array}\,\leq\,0\,.\label{cglmp}\ea It is conjectured that all
non-trivial facets are equivalent to the CGLMP inequality for all
$d$. Anyway, our work is independent of the truth of this
conjecture: we are going to study the possibility of secret key
extraction for non-local distributions which lie above a CGLMP
facet, irrespective of whether there exist inequivalent facets or
not.

The {\em no-signalling polytope} appears to have a richer
structure than in the case $d=2$. All the extremal non-local
points are generalizations of the PR-box \cite{barrett}. We are
interested in those that lie above our representative CGLMP facet.
The highest violation of CGLMP is provided by the extremal point
\ba PR_{2,d} &=& \frac{1}{d}\,\delta(b-a=xy)\,, \label{prd}\ea
whose corresponding array is \ba PR_{2,d}&=& \frac{1}{d}\,
\begin{array}{c|cccc|cccc|} A\setminus B& 1&1&\ldots &1& 1& 1&\ldots
&1\\\hline 1&  1&0&\ldots &0& 1&0&\ldots &0\\ 1& 0&1&\ldots &0&
0&1&\ldots &0\\ \vdots & \vdots & &\ddots & \vdots& \vdots &
&\ddots & \vdots\\ 1& 0&0&\ldots &1& 0&0&\ldots &1\\\hline 1&
1&0&\ldots &0& 0&1&\ldots &0\\ 1&  0&1&\ldots &0& 0&0&\ddots &\vdots\\
\vdots & \vdots & &\ddots & \vdots& \vdots &\ddots &\ddots& 1\\ 1&
0&0&\ldots &1& 0&0&\ldots &0\\\hline
\end{array}\;. \label{pr2d}\ea Its violation of the inequality can be rapidly calculated by
a term-by-term multiplication (a formal scalar product) of the two
arrays (\ref{cglmp}) and (\ref{pr2d}), yielding \ba
\left<I_d,PR_{2,d}\right> &=& \frac{d-1}{d}\,. \ea However,
$PR_{2,d}$ is not the only non-local extremal point which lies
above a CGLMP facet: in fact, for all $d'<d$, there is at least
one $PR_{2,d'}$ above the facet. For instance, a possible version
of $PR_{2,2}\equiv PR$ reads (boldface $\mathbf{0}$ standing for
matrices filled with zeros) \ba PR&=& \frac{1}{2}\,
\begin{array}{c|cc|cc|} A\setminus B& \begin{array}{cc} 1&1\end{array} & \mathbf{0} &
\begin{array}{cc} 1&1\end{array} & \mathbf{0}\\\hline
\begin{array}{cc} 1\\1\end{array} & \begin{array}{cc} 1&0\\0& 1\end{array} & \mathbf{0} &
\begin{array}{cc} 1&0\\0& 1\end{array} & \mathbf{0}\\
\mathbf{0} & \mathbf{0} & \mathbf{0} & \mathbf{0} &
\mathbf{0}\\\hline
\begin{array}{cc} 1\\1\end{array} &
\begin{array}{cc} 1&0\\0& 1\end{array} & \mathbf{0} & \begin{array}{cc} 0&1\\1&
0\end{array}
 & \mathbf{0}\\
\mathbf{0} & \mathbf{0} & \mathbf{0} & \mathbf{0} &
\mathbf{0}\\\hline\end{array}\;,\ea whence a violation
$\left<I_d,PR\right> = \frac{1}{2}$. For $d=3$, we shall give
below (\ref{secd3}) some additional elements on the structure of
the no-signalling polytope.

The boundaries of the {\em quantum region} are basically unknown
to date; it is not even clear whether they coincide with all
possible results of measurements on two-qutrits states.

\subsubsection{A slice in the non-local region}
\label{secdhyp}

As we said, a no-signalling probability distribution is
characterized by $4d(d-1)$ parameters. However, when one reviews
the results obtained for the CGLMP inequality in the context of
quantum physics (see Appendix \ref{appsettings} for all details),
one finds that the probability distributions associated to the
optimal settings belong to a very symmetric family. Specifically,
these distributions are such that (i) for fixed inputs $x$ and
$y$, $P(a,b|x,y)$ depends only on $\Delta=a-b$; and (ii) the
probabilities for the different inputs are related as
$P(\Delta|0,0)=P(-\Delta|0,1) =P(-\Delta|1,0)=P(\Delta-1|1,1)$.
Compactly: \ba
P(a,b=a-\Delta|x,y)&=&\frac{1}{d}\,p_{f}\label{slice}\ea with
$f=(-1)^{x+y}\Delta+xy$ and $p_f= \sum_{a}P(a,b=a-f|0,0)$. The
corresponding array is \ba P&=& \frac{1}{d}\,
\begin{array}{c|cccc|cccc|} A\setminus B& 1&1&\ldots &1& 1& 1&\ldots
&1\\\hline 1&  p_0&p_{-1}&\ldots &p_2& p_0&p_1&\ldots &p_{-2}\\
1& p_1&p_0&\ldots &p_{3}& p_{-1}&p_0&\ldots &p_{-3}\\ \vdots &
\vdots & &\ddots & \vdots& \vdots & &\ddots & \vdots\\ 1&
p_{-2}&p_{-3}&\ldots &p_0& p_2&p_3&\ldots &p_0\\\hline 1&
p_0&p_{1}&\ldots &p_{-2}& p_{1}&p_0&\ldots &p_{3}\\
1&  p_{-1}&p_0&\ldots &p_{-3}& p_2&p_1&\ldots &p_4\\
\vdots & \vdots & &\ddots & \vdots& \vdots & &\ddots& \vdots\\
1& p_2&p_3&\ldots &p_0& p_{-1}&p_{-2}&\ldots &p_1\\\hline
\end{array}\;.\label{parray}\ea This family defines a {\em slice} in the no-signalling
polytope. Note that all the marginals are equal, that is,
$P(a|x)=P(b|y)=\frac{1}{d}$. Moreover, the $d$ numbers $p_f$
define uniquely and completely a point $P$ in the slice; thus,
given the constraint $\sum_fp_f=1$, the slice defined by
(\ref{slice}) is $(d-1)$-dimensional. A single extremal non-local
point belongs to the slice, namely $PR_{2,d}$, obtained by setting
$p_0=1$ (\ref{prd}); in fact, none of the $PR_{d'}$ with $d'<d$
has the correct marginals.

As it happened for the isotropic distributions for $d=2$, there
exists a depolarization procedure that maps any probability
distribution onto this slice by local operations and public
communication, while keeping the violation $\left<I_d,P\right>$
constant. The procedure is given in Appendix \ref{appdepolarize}.
As a consequence, Eve's optimal individual eavesdropping, for a
fixed value of the violation of the inequality, consists in
distributing a point in the slice.

\subsection{Cryptography}

\subsubsection{The protocol}

We suppose from the beginning $p(x=i)=p(y=j)=\demi$. The protocol
is the analog of the CHSH protocol described in \ref{chshproto}
above. When Alice announces $x=1$ and Bob has measured $y=1$, Bob
corrects his dit according to $b\rightarrow b-1$. In other words,
the pseudo-sifting implements $\Delta\rightarrow \Delta-xy$. The
Alice-Bob distribution after pseudo-sifting, averaged on Bob's
settings, becomes independent of $x$ (as in the case of isotropic
distribution for $d=2$): \ba P(a,a-\Delta|x)&=&
\frac{1}{d}\,\sum_{y=0,1}p_{(-1)^{x+y}\Delta}\,=\,\frac{p_\Delta +
p_{-\Delta}}{2d}\,. \ea In a protocol with $d$-dimensional
outcomes, Alice and Bob can estimate not just one, but several
error rates, one for each value of $\Delta$. We have just found
that these error rates exhibit the symmetry \ba
e_{AB}(\Delta)\,=\,e_{AB}(-\Delta)&=&\frac{p_\Delta +
p_{-\Delta}}{2}\,.\label{errors}\ea As in the case of bits, we
think of Eve as sending either a local or a non-local probability
distribution. Let's discuss in some detail the points which lie on
and above a CGLMP facet.

\subsubsection{Eve's strategy: local points}
\label{ssloc}

To understand what follows, we don't need a full characterization
of the deterministic strategies that saturate the CGLMP
inequality. Some facts are however worth noting; the proof of
these statements and some other features are given in Appendix
\ref{apptechn}.

The first fact is that, for $d>2$, the number of deterministic
points on the CGLMP facet is strictly larger than $D=4d(d-1)$, the
dimension of the local and the no-signalling polytope. This
implies that, for some points on the facet, several decomposition
as a convex combination of extremal points are possible.

The second fact is that no extremal deterministic strategy belongs
to the slice (\ref{slice}): to see it, just recall that the
marginals in the slice are completely random. Since we require the
final distribution to belong to the slice, Eve must manage to send
deterministic strategies with the suitable probabilities. As a
consequence of the previous remark, at least one local point on
the slice can be obtained by several different decompositions on
extremal points: we'll have to choose the decomposition that
optimizes Eve's information.

As a third fact, we elaborate on the same idea that lead to the
uncertainty relations in paragraph \ref{sec2unc}. We know that all
deterministic strategies are not equally interesting for Eve, in
fact, two kind of local points are of special interest for her:
(i) those for which $b(0)=b(1)$, because Eve knows Bob's symbol
when Alice announces $x=0$, and which we denote by the set
${\cal{L}}_0$; and (ii) those for which $b(0)=b(1)-1$, because Eve
knows Bob's symbol when Alice announces $x=1$, and which we denote
by the set ${\cal{L}}_1$. In all the other cases, Eve does not
learn Bob's symbol with certainty. Now, in the complexity of the
list of deterministic points on  the CGLMP facet, a remarkable
feature appears:
\begin{itemize}
\item There are exactly $d^2$ points in ${\cal{L}}_0$, namely
those for which $a(0)=b(0)=b(1)$ and $a(1)$ can take any value. In
other words, there are no points on the CGLMP facet such that
$b(0)=b(1)$ but $a(0)$ is different from this value: whenever Eve
learns Bob's symbol for $x=0$, Alice and Bob make no error for
$x=0$.

\item There are exactly $d^2$ points in ${\cal{L}}_1$, namely
those for which $a(1)=b(0)=b(1)-1$ and $a(0)$ can take any value.
This has a similar interpretation as the statement above, in the
case $x=1$.
\end{itemize}
Now, since the error rate Alice-Bob depends only on $P(a,b|x,y)$,
and not on the particular decomposition chosen by Eve to realize
this distribution, it is obvious that Eve's interest lies in
distributing local points that belong to ${\cal{L}}\equiv
{\cal{L}}_0\cup{\cal{L}}_1$ as often as possible. For $d=3$, we
shall prove that she can prepare any point in the slice by
distributing only these kind of local points. Finally, we want to
introduce a further distinction within ${\cal{L}}$, which appears
explicitly in the study of $d=3$ but may play a more general role.
We shall call ${\cal{L}}^3$ the subset of ${\cal{L}}$, whose
points satisfy three out of the four relations $a(0)=b(0)$,
$a(0)=b(1)$, $a(1)=b(0)$ and $a(1)=b(1)-1$; the complementary set,
containing the points that satisfy two of the relations
($a(0)=b(0)=b(1)$ or $a(1)=b(0)=b(1)-1$), is written
${\cal{L}}^2$.

\subsubsection{Eve's strategy: non-local point}

We have said that, among the extremal non-local points which lie
above the CGLMP facet, the only one on the slice (\ref{slice}) is
$PR_{2,d}$. However, it may be the case that mixtures of other
extremal non-local points lie as well in the slice. For $d=3$,
this is not the case (see Appendix \ref{apptechn3}), but we have
not been able to generalize this statement. In this study, we
suppose tentatively that Eve sends a unique non-local strategy,
namely $PR_{2,d}$. Under this assumption, we can define $p_{NL}$
as the probability that Eve sends $PR_{2,d}$. To find the
expression of $p_{NL}$, we notice that $\left<I_d,PR_{2,d}\right>
=\frac{d-1}{d}$ should correspond to $p_{NL}=1$, and that
$\left<I_d,L\right> =0$ for all local points on the CGLMP facet,
should correspond to $p_{NL}=0$. Moreover, $p_{NL}$ measures the
geometrical distance from the facet and is therefore an affine
function of the violation of CGLMP. Thus for a generic
distribution $P$ of the form (\ref{parray}) we have \ba p_{NL}&=&
\frac{d}{d-1}\, \left<I_d,P\right>\,=\nonumber\\&=&
-2+\sum_{\Delta=0}^{d-1}\left(1-\frac{\Delta}{d-1}\right)\,\left[3p_{-\Delta}
-p_{\Delta+1}\right]\,.\label{pnl}\ea

Now we can present the results for the possibility of extracting a
secret key, starting from a detailed study of the case $d=3$
(\ref{secd3}), then generalizing some results for arbitrary $d$
(\ref{secdres}).

\subsection{Secret key extraction: $d=3$}
\label{secd3}

\subsubsection{The slice of the polytope}

\begin{figure*}[t]
\includegraphics[width=8cm]{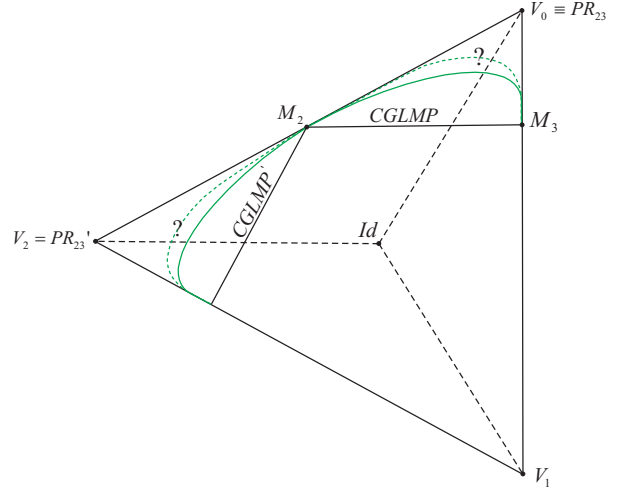}
\caption{The slice (\ref{slice}) of the no-signalling polytope,
for $d=3$. As explained in the text, the full extent of the
quantum region is not known, it is represented by the dotted line
with question marks. The full line is the part of the quantum
region we can certainly reach and that we consider in this paper.
See text for all the other details.}\label{fig31}
\end{figure*}

The slice (\ref{slice}) is 2-dimensional for $d=3$, we choose
$p_0$ and $p_1$ as free parameters; this gives $p_2=1-p_0-p_1$ and
\ba p_{NL}&=&2(p_0-p_1)-1\,.\ea The full slice has a form of an
equilateral triangle (Fig.~\ref{fig31}), whose vertices $V_\Delta$
are defined by $p_\Delta=1$. As mentioned, $V_0=PR_{2,3}$. The
vertex $V_2$ is also a $PR_{2,3}$, the one defined by
$b-a=\bar{x}\bar{y}+1$ with $\bar{z}=1-z$. On the contrary, $V_1$
a mixture of deterministic strategies. The middle of the triangle,
$p_0=p_1=p_2=\frac{1}{3}$, is the completely random strategy
(obtained e.g. when measuring the maximally mixed quantum state,
the "identity").

\begin{figure*}
\includegraphics[width=8cm]{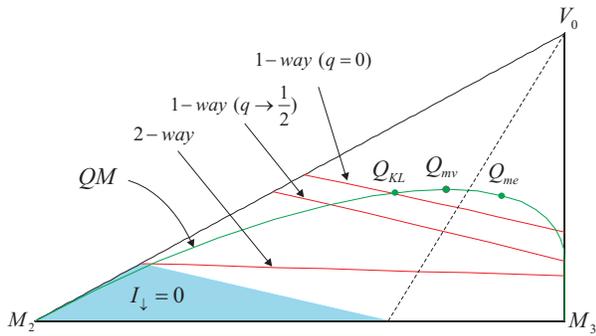}
\caption{Zoom of Fig.~\ref{fig31} on the non-local region close to
$V_0$. For clarity, only the part of the quantum region that we
consider is represented here. The transverse lines define the
limits down to which secrecy can be extracted for one-way
post-processing (without and with pre-processing) and for two-way
post-processing without pre-processing. In the shaded region, the
intrinsic information $I_{\downarrow}$ is zero. We stress that
this figure is an exact plot, not just an "artist view". See text
for the other details.}\label{fig32}
\end{figure*}

We are going to focus on the non-local region close to $V_0$
(Fig.~\ref{fig32}). The intersection with the CGLMP facet is the
segment $p_0-p_1=\demi$, whose ends are the points labelled $M_2$
($p_0=\demi$, $p_1=0$) and $M_3$ ($p_0=\frac{3}{4}$,
$p_1=\frac{1}{4}$). The decompositions of these mixtures on the
extremal deterministic strategies are \ba M_2=\sum_{L\in
{\cal{L}}^2}\,\frac{1}{6}\,L&\;,\;\;& M_3=\sum_{L\in
{\cal{L}}^3}\,\frac{1}{12}\,L \label{decomp}\ea where the sets of
local points ${\cal{L}}^2$ and ${\cal{L}}^3$ have been defined
above. In fact, the decomposition of $M_3$ is unique; conversely,
$M_2$ can be decomposed in an infinity of ways (see Appendix
\ref{apptechn3}), but all the others involve also the points that
don't belong to ${\cal{L}}$ and are therefore sub-optimal for Eve.

The quantum-mechanical studies (see Appendix \ref{appsettings} for
more details) have singled out two non-local probability
distributions in this region. The first one corresponds to the
maximal violation of CGLMP using two qutrits, $p_{NL}\approx
0.4574$: it is noted $Q_{mv}$ and is defined by (\ref{probsgamma})
with $\gamma=\frac{\sqrt{11}-\sqrt{3}}{2}$. The second one
corresponds to the highest violation achievable with the maximally
entangled state of two qutrits, $p_{NL}\approx 0.4365$: it is
noted $Q_{me}$ and is defined by (\ref{probsgamma}) with
$\gamma=1$.

\subsubsection{One-way classical post-processing}

To write down the table for the correlations Alice-Bob-Eve, one
needs to list explicitly the deterministic points that saturate
CGLMP and the corresponding information Eve can extract. This is
done in Appendix \ref{apptechn3}. The result is Table
\ref{tabled3}. It can be verified easily that all the
probabilities in a row/column sum up to $\frac{1}{3}$; moreover,
\ba e_{AB}(+1)\,=\,e_{AB}(-1)&=&\frac{1-p_0}{2} \label{eab3}\ea as
expected from (\ref{errors}). We have introduced the symbol $?_2$
to describe the situation where Eve is uncertain on Bob's symbol,
but only among two possibilities: this is clearly the case
whenever the uncertainty derives from a deterministic strategy. In
all that follows, information is quantified in trits, and we write
$h\big([v_1,v_2,v_3]\big)=-\sum_j v_j\log_3 v_j$.

In the absence of pre-processing, Eve has no information with
probability $p_{NL}$, full information with probability
$\frac{p_L}{2}$, and information
$1-h\big([1/2,1/2,0]\big)=1-\log_32$ with probability
$\frac{p_L}{2}$. Therefore the estimate for the CK bound is \ba
R_{CK}(q=0)\,\geq\, r_{CK}&=&1-h\left(\big[p_0,\frac{1-p_0}{2},
\frac{1-p_0}{2}\big]\right) \nonumber\\ &&- p_L\left(1-\demi
\log_32 \right)\,.\ea The curve $r_{CK}(q=0)=0$ is shown in
Fig.~\ref{fig32}, it clearly cuts the quantum region.

A natural question is, which is the point that maximizes
$r_{CK}(q=0)$ under the requirement that the correlations should
belong to the quantum region. In the slice under consideration, we
find a rate $r_{max}(q=0)\approx 0.09$ trits $\approx 0.144$ bits
for the correlations defined by $p_0=0.8286$, $p_1=0.1093$. These
correlations can be obtained by measuring the quantum state \ba
\ket{\psi(\gamma)}&=&
\frac{1}{\sqrt{2+\gamma^2}}\,\big(\ket{00}+\gamma
\ket{11}+\ket{22}\big)\label{psigamma} \ea for $\gamma\approx
0.9875$. This state is close to, but certainly different from, the
maximally entangled state. Thus, the secret key rate exhibits the
same form of anomaly as all the other measures of non-locality
known to date \cite{methot}: maximal non-locality is obtained with
non-maximally entangled states.

We consider now Bob's pre-processing. For one-way post-processing,
dit-wise pre-processing is already optimal. A priori, one can
define two different flipping probabilities $q_{+1}$ and $q_{-1}$,
associated respectively to $b\rightarrow b+1$ and $b\rightarrow
b+2$. But it turns out by inspection that the optimal is always
obtained for $q_{+1}=q_{-1}=q$, so we write down directly this
case. From (\ref{eab3}) it is clear that after pre-processing \ba
e_{AB}'(+1)=e_{AB}'(-1)&\equiv&
\frac{e'}{2}\,=\,\frac{1-p_0}{2}+q\,\frac{3p_0-1}{2} \ea whence
\ba I(A:B')&=&1-h\left([1-e',\frac{e'}{2},\frac{e'}{2}]\right)\,.
\label{iab3} \ea Eve's information is computed by recalling that,
for any local point she sends out, before pre-processing (i) for
one value of $x$, she knows perfectly Bob's symbol $b$; (ii) for
the other value of $x$, she hesitates between two values of $b$.
Pre-processing leaves $b$ unchanged with probability $1-2q$, and
sends it to $b\pm 1$ with probability $q$ each. Therefore, in case
(i), Eve's information is lowered from 1 to $1-h([1-2q,q,q])\equiv
1-h_1(q)$; in case (ii), Eve's information is lowered from
$1-h([\demi,\demi,0])$ to
$1-h([\frac{1-q}{2},\frac{1-q}{2},q])\equiv 1-h_2(q)$. Since each
case is equiprobable, \ba
I(E:B')&=&1-\demi\left[h_1(q)+h_2(q)\right]\,. \label{ibe3}\ea
From (\ref{iab3}) and (\ref{ibe3}), we can compute $r_{CK}$ by
optimizing the value of $q$. We did the optimization numerically.
The improvement due to pre-processing is clear in
Fig.~\ref{fig32}.

\subsubsection{Two-way classical post-processing}

We have also studied the possibility of extracting a secret key
from the correlations of Table \ref{tabled3} using AD (without
pre-processing). Alice selects $N$ of her symbols that are
identical, Bob accepts if and only if his corresponding symbols
are also identical. The probability that Bob accepts is $p_0^{N}+
[e_{AB}(+1)]^N+[e_{AB}(-1)]^N$, and consequently \ba
\tilde{e}_{AB}(\pm 1)&=&
\frac{\left(\frac{1-p_0}{2}\right)^N}{p_0^N+2\left(\frac{1-p_0}{2}\right)^N}
\approx \left(\frac{1-p_0}{2p_0}\right)^N\,.\ea As in the case
$d=2$, Eve has to make a random guess if and only if she has sent
$PR_{2,3}$ for all the $N$ instances: \ba \tilde{e}_{E}(\pm
1)&\gtrsim& \frac{1}{3}\, \left(\frac{p_{NL}}{p_0}\right)^N\,.\ea
Thus, a secret key can be extracted using AD as long as
$p_{NL}>\frac{1-p_0}{2}$, that is as long as \ba
5p_{0}&>&4p_1+3\,. \ea The limiting curve is also plotted in
Fig.~\ref{fig32}. Its extremal points are $p_{NL}>\frac{1}{5}$ for
$p_1=0$ (the same value as obtained for $d=2$) and
$p_{NL}>\frac{1}{9}$ for $p_2=0$.

\subsubsection{Intrinsic information}

It is straightforward to generalize the map used above in the
computation of the intrinsic information to the $d=3$ case.
Looking at table \ref{tabled3}, one has to map all the symbols
$(i,?_2)$ into $(?,?)$, where $i=0,1,2$. The obtained conditional
mutual information reads \ba \label{upintr}
    I(A:B|\bar E)&=&P(?,?)\,\times\nonumber\\ &&\left[1-h\left(\frac{2p_0-p_L}{2-p_L},
    \frac{1-p_0}{2-p_L},\frac{1-p_0}{2-p_L}\right)\right]\, .
\ea This is of course an upper bound to the intrinsic information,
since the employed map may not be the optimal one. Contrary to
what happens in the $d=2$ case, this quantity vanishes for some
points inside the region of Bell violation! Indeed,
Eq.~(\ref{upintr}) is zero on the line $5p_0-2p_1-3=0$; by
changing slightly Eve's map (specifically, she applies the map
above only with a suitable probability and makes nothing in the
other cases), it can be verified that the intrinsic information is
zero also below the line, that is for \ba
    5p_0-2p_1-3&\leq &0\,.
\ea This region overlaps with the non-local region
(Fig.~\ref{fig32}).

\subsection{Secret key extraction: generic $d$}
\label{secdres}

For generic $d$, we want to prove that secrecy can be generated
using quantum states. The statistics Alice-Bob can be computed
using quantum mechanics, in particular the error rates
$e_{AB}(\Delta)$ of Eq.~(\ref{errors}). The question is, how to
estimate Eve's information: to compute this quantity exactly, one
must describe the points in the CGLMP facet in some detail.
However, some interesting bound can be derived from what we have
already said and the intuition developed in the study of $d=3$.

Consider first {\em one-way post-processing}: the discussion of
paragraph \ref{ssloc} implies the bound \ba I(B:E)&\leq&
I_{E}\,\equiv\,\frac{p_L}{2}\,+\,\frac{p_L}{2}\left(1-\log_d
2\right)\,. \ea The bound is reached if and only if Eve
distributes strategies that belong to ${\cal{L}}$, as it happened
to be always possible for $d=3$. Moreover, this bound can also be
computed from the Alice-Bob distribution only assuming
(\ref{pnl}). Consequently we can estimate \ba R_{CK}(q=0)&\geq&
{r}\equiv 1-h\left(\{e_{AB}(\Delta)\}_{\Delta}\right)-I_E
\label{d1w}\ea with $h$ the Shannon entropy measured in dits. We
have studied the r.h.s. numerically for $d\leq 10$, for
correlations in the quantum region obtained from states that are
Schmidt-diagonal in the computational basis,
$\ket{\psi}=\sum_{k=0}^{d-1}c_k\ket{k\,k}$. The general features
that emerge are:
\begin{itemize}
\item The maximal value of ${\cal{R}}$ achievable in the quantum
region increases with $d$, reaching up to ${\cal{R}}\approx 0.692$
bits for $d=10$.

\item The quantum state corresponding to the maximal value of
${\cal{R}}$ is always such that $c_k=c_{d-1-k}$. It seems that the
overlap $\eta$ of this state with the maximally entangled one
decreases with $d$, but the decrease is very slow (we have
$\eta=1$ for $d=2$, and for $d=10$ we still have $\eta\gtrsim
0.998$).
\end{itemize}

A similar simple approach can be found to explore the
possibilities of {\em two-way post-processing}. We have \ba
\tilde{e}_{AB}(\Delta)&\sim&[e_{AB}(\Delta)]^N\,\leq\,
\left[\max_{\Delta=1,...,d-1}e_{AB}(\Delta)\right]^N\ea and Eve's
error is $\tilde{e}_E\sim \frac{1}{d}\,p_{NL}^N$. Consequently, AD
will certainly work for \ba p_{NL}&> &
\max_{\Delta=1,...,d-1}e_{AB}(\Delta)\,. \label{d2w}\ea All the
quantities in this relation can be computed from the Alice-Bob
correlations alone. As before, we have studied this condition
numerically, for $d\leq 100$. This time, we have concentrated on
correlations of the form $P\,=\,w\,P_{me}\,+\,\frac{(1-w)}{d}$,
where $P_{me}$ are the correlations obtained when measuring the
maximally entangled state (this is of course a completely
arbitrary choice, but seems interesting from the point of view of
quantum physics). One observes that, as expected, the use of
two-way post-processing significantly decreases the value
$p_{NL}(0)$ of $p_{NL}$ for which no secrecy can be extracted.
Moreover, $p_{NL}(0)$ decreases when $d$ increases, but very
slowly; so slowly in fact, that it cannot be guessed from the
numerical results whether ultimately $p_{NL}(0)\rightarrow 0$ for
$d\rightarrow \infty$.

In summary, we have obtained a few results for generic $d$. In
spite of a large number of assumptions and approximations (not
least the choice of the protocol), we can conjecture that secrecy
can be extracted from quantum non-local correlations for any $d$;
and more precisely, that the amount of extractable secrecy
increases with increasing $d$.

\section{Conclusions and Perspectives}
\label{secpersp}

In conclusion, we have presented a first approach to a
device-independent security proof for cryptography, expanding and
generalizing the work of Ref.~\cite{prl}. Under the assumption of
individual attacks, we have proved that a secret key can be
extracted from some no-signalling probability distributions, using
only the very fact that they violate a Bell-type inequality and
cannot therefore originate from shared randomness. In particular,
noisy quantum states can be used to distribute correlations, that
are non-local enough to contain distillable secrecy: so our result
is also of practical interest.

We'd like to finish by raising some of the questions and
perspectives that are opened by this work.

\begin{itemize}

\item A first objective is to extend our analysis beyond the
assumption of individual attacks, proving ultimately the security
against the most general attacks by an eavesdropper limited by
no-signalling. A first step in this direction has been recently
derived \cite{masaneswinter}.

\item One can make a step further: can one make a
device-independent proof of security against an eavesdropper which
would be limited by quantum physics? On the side of Alice and Bob,
non-locality should still be the physical basis for security,
because there exist no other entanglement witness which works
independently of the dimension of the Hilbert space. On the side
of Eve, the requirement that she must respect quantum physics is a
limitation, compared to power we gave her in this paper; so one
can hope to obtain a device-independent proof with better bounds.

\item In this paper, we have defined protocols which look as
"natural" for the CHSH and the CGLMP inequalities. But there is no
claim of optimality. In fact, it is not even proved that the
pseudo-sifting that we have used is the best way of extracting
secrecy from the raw correlations of CHSH-like measurements (Table
\ref{tableraw}). Other protocols may be better suited for
cryptographic tasks, as discussed in Ref.~\cite{amp06}.

\item A particular consequence of the previous item is worth
mentioning in itself. On the one hand, it has been proved that all
non-local probability distributions have positive intrinsic
information \cite{generic}. On the other hand, as mentioned
several times in this paper, we have not been able to find an
explicit procedure for extracting a secret key in the whole
non-local region. This means, either that a better procedure does
exist, or that non-local distributions close to the local limit
provide examples of {\em bipartite bound information}
\cite{GW,multbound}.

\item A technical open point, which we mentioned and would be very
meaningful for the present studies, is the characterization of the
quantum region in probability space for a given number of inputs
and outcomes.

\end{itemize}

\section*{Acknowledgements}

We thank Stefano Pironio, Sandu Popescu and Renato Renner for
discussion. This work has been supported by the European
Commission, under the Integrated Project Qubit Applications (QAP)
funded by the IST directorate as Contract Number 015848, and the
Spanish MEC, under a Ram\'on y Cajal grant. We acknowledge also
financial support from the Swiss NCCR "Quantum Photonics".

\begin{appendix}

\section{Lower bound for a quantum implementation of the CHSH protocol}\label{appcrypto}

In this appendix, we study the security of the CHSH protocol in
the standard scenario where Eve is limited by the quantum
formalism, and Alice and Bob have a perfect knowledge on their
quantum devices. More precisely, Alice and Bob know their Hilbert
spaces are two-dimensional and they apply the spin measurements
that produce the largest Bell violation for the noiseless state
$\ket{\Phi^+}=\frac{1}{\sqrt 2}(\ket{00}+\ket{11})$. For instance,
Alice and Bob measure in the $xz$ plane, their spin measurement
being defined by the angle $\theta$ with the $z$ axis on the
Poincar\'e sphere. Alice measures in the $\theta=\pi/2$ and
$\theta=0$ bases, corresponding to $x=0,1$ respectively, while Bob
does it in the $\theta=\pi/4,3\pi/4$ directions, corresponding to
$y=0,1$.

As shown in Refs \cite{KGR,Renner}, the bound for security against
the most general attacks ("unconditional security") can be
computed by focusing on "collective attacks", where Eve prepares
the same two-qubit state $\rho_{AB}$ on all instances, but is
allowed to make a coherent measurement of her ancillae after error
correction and privacy amplification.

By inspection, or by using the formalism developed in Ref.
\cite{KGR}, it can be proved that Eve's optimal strategy uses a
Bell-diagonal state of the form
\begin{equation}\label{qattack}
    \rho_{AB}=\lambda_1 \Phi^+ + \lambda_2(\Phi^-+\Psi^+)+\lambda_4\Psi^- ,
\end{equation}
where $\Phi^\pm$ and $\Psi^\pm$ denote the projectors onto the
Bell basis
\begin{eqnarray}
  \ket{\Phi^\pm} &=& \frac{1}{\sqrt 2}(\ket{00}\pm\ket{11}) \nonumber\\
  \ket{\Psi^\pm} &=& \frac{1}{\sqrt 2}(\ket{01}\pm\ket{10}) .
\end{eqnarray}
By assumption, Eve holds a purification of each pair: before any
measurement, the quantum correlations among Alice, Bob and Eve are
described by the pure state $\ket{\psi_{ABE}}^{\otimes N}$ where
$\rho_{AB}=\tr_E\ket{\psi}\bra{\psi}_{ABE}$.

In the CHSH protocol, Alice and Bob's bases do not perfectly
overlap: their outcomes are therefore not perfectly correlated
even in the case $\lambda_1=1$ (perfect channel, no Eve):
actually, the quantum bit error rate (QBER) in this case is $Q_0=
\sin^2(\pi/8)=\demi\left(1-\frac{1}{\sqrt{2}}\right)$. For the
same channel, the BB84 protocol has zero QBER. In the light of
this, the meaningful parameter to compare the two protocols should
not be the QBER, but a measure of the quality of the channel. We
use the {\em disturbance}, that is the probability that
measurement outcomes in the same basis agree: in our case,
$D=\bra{+z,-z}\rho_{AB}\ket{+z,-z}+\bra{-z,+z}\rho_{AB}\ket{-z,+z}=\lambda_2+\lambda_4$.

Now, Alice and Bob measure their local systems, while Eve keeps
her quantum state. In this scenario, a lower bound to the key rate
distillable using one-way communication protocols has been
obtained in \cite{DW},
\begin{equation}\label{dwbound}
    R^\rightarrow\geq R_{DW}=I(A:B)-\chi(B:E) .
\end{equation}
Here, $I(A:B)$ denotes the standard mutual information between
Alice and Bob's classical outcomes, while $\chi(B:E)$ is the
Holevo quantity for the effective channel between Bob and Eve.
Indeed, Bob's measurement outcome prepares a quantum state on
Eve's site (see \cite{DW} for more details). Contrary to the more
standard situation, Eve does not know which measurement Bob has
applied, so she has to sum over the two possibilities. Her states
read, up to normalization,
\begin{equation}
    \rho_E^i=\tr_{AB}\Big[\one\otimes\left(\ket{i}\bra{i}_{\pi/4}+
    \ket{i}\bra{i}_{3\pi/4}\right)\otimes \one
    \ket{\psi}\bra{\psi}_{ABE}\Big] ,
\end{equation}
where $i=0,1$ and $\ket{i}_\theta$ denote the basis elements in
the direction specified by $\theta$, as above. It is
straightforward to see that for the CHSH protocol
\begin{equation}
    I(A:B)=1-h\left(\frac{1+(\lambda_1-\lambda_4)/\sqrt 2}{2}\right) .
\end{equation}
The computation of
$\chi(B:E)=S(\rho_E)-(S(\rho_E^0)+S(\rho_E^1))/2$, where $S(\rho)$
denotes the von Neumann entropy for a state $\rho$, is slightly
more involved. However, after some patient algebra one can see
that the maximum of this quantity is obtained, for fixed
disturbance, when
\begin{equation}\label{optatt}
    \lambda_1=(1-D)^2\quad \lambda_2=D(1-D)\quad
    \lambda_4=D^2 ,
\end{equation}
which defines Eve's optimal attack. Not surprisingly, this attack
corresponds to a phase covariant cloning machine (see for instance
\cite{Cerf}), that optimally clones all the states in the $xz$
plane. This attack is also optimal for the standard BB84 protocol.

The obtained critical disturbance for CHSH is $D\approx 12\%$.
This is larger than the well-known Shor-Preskill bound $D\approx
11\%$ for security of BB84 \cite{SP}. This bound, however, has
recently been improved by allowing any of the parties, say Alice,
to introduce some {\em pre-processing} of her outcome before the
reconciliation \cite{KGR}. Alice then flips her bit with
probability $q$. This local noise worsens the correlations between
Alice and Bob but it deteriorates in a stronger way the
correlations between Alice and Eve. For any value of the
disturbance there exists an optimal pre-processing $q(D)$,
depending on the protocol, which maximizes the key rate. This
explains the improvement on the critical disturbance that moves up
to $D\approx 12.4\%$ both for BB84 and for the CHSH protocol
described here.

Actually, the close relation between the CHSH protocol and the
BB84 protocol is made clear by this pre-processing. Let $Q_B\equiv
D$ and $q_B$ be respectively the QBER and the pre-processing rate
for BB84; and $Q_C$ and $q_C$ denote the same quantities for the
CHSH protocol. Note first that the channel defined in
(\ref{optatt}) induces a QBER $Q_B=D$ in BB84, and a QBER
$Q_C=Q_0+\frac{D}{\sqrt{2}}$ for the CHSH protocol; whence \ba
Q_B&=&\sqrt{2}(Q_C-Q_0)\,.\label{relqber}\ea It can then be shown
that $R_{DW}^{CHSH}(Q_C,q_C)= R_{DW}^{BB84}(Q_B,q_B)$ when the
QBERs are related as (\ref{relqber}) and when \ba
q_B&=&Q_0+\frac{q_C}{\sqrt{2}}\,.\label{relq}\ea These two
relations imply \ba Q'&=&Q_C(1-q_C)+(1-Q_C)q_C\nonumber\\ &=&
Q_B(1-q_B)+(1-Q_B)q_B\,.\label{relqprime}\ea

Consider first for clarity the case $q_C=0$: the error in CHSH due
to the non-perfect overlap of the bases can be attributed to the
application of a pre-processing $q_B=Q_0$ onto the correlations
obtained with perfectly overlapping bases (indeed, the errors
$Q_0$ are intrinsic to the protocol, and Eve cannot gain anything
from them). In general, the rates obtained for the CHSH protocol
in the standard quantum scenario coincide with those derived for
the BB84 protocol when the pre-processing is optimized under the
constraint $q_B\geq Q_0$. If we now compare $R_{DW}^{CHSH}(Q_C)$
and $R_{DW}^{BB84}(Q_B)$ for a fixed value of $D$ [that is,
(\ref{relqber}) holds] and choosing the optimal pre-processing in
each case, we find the following. For small error rates, the
optimal pre-processing on BB84 is smaller than $Q_0$; in other
words, even for $q_C=0$ CHSH corresponds to BB84 with an excessive
pre-processing, whence $R_{DW}^{CHSH}(Q_C)<R_{DW}^{BB84}(Q_B)$.
The optimal pre-processing on BB84 becomes equal to $Q_0$ for
$D\approx 11.7\%$; from this point on, the optimal $q_C$ is larger
than zero, and the rates for the two protocols become identical:
$R_{DW}^{CHSH}(Q_C)=R_{DW}^{BB84}(Q_B)$. In particular, as
announced, both become zero for $D\approx 12.4\%$.

\section{The CGLMP inequality in quantum physics}\label{appsettings}

The CGLMP inequalities \cite{cglmp} have been the object of
several studies in the context of quantum physics. Here we
summarize the results without any proof.

One unexpected features of CGLMP is the fact that the maximal
violation is not reached by measurements on the maximally
entangled state \cite{adgl02}. Also unexpected is the fact that
the settings that maximize the violation are the same for a wide
class of states (including the maximal entangled one and the one
which gives the maximal violation). These are the settings we
consider here. We label them $A_0$ and $A_1$ for Alice, $B_0$ and
$B_1$ for Bob: \ba
A_x\equiv\left\{\Psi_{x}(a)\right\}_{a=0}^{d-1}&\;,\;&
\Psi_{x}(a)=
\sum_{k=0}^{d-1}\frac{e^{i\frac{2\pi}{d}ak}}{\sqrt{d}}
\left(e^{ik\phi_x}\ket{k}\right)\,,\\
B_y\equiv\left\{\Phi_{y}(b)\right\}_{b=0}^{d-1}&\;,\;&
\Phi_{y}(b)=
\sum_{k=0}^{d-1}\frac{e^{-i\frac{2\pi}{d}bk}}{\sqrt{d}}
\left(e^{ik\theta_y}\ket{k}\right)\,.\ea In operational terms,
both Alice and Bob apply first global phases in the computational
basis, then make a quantum Fourier transform (Bob makes the
inverse as Alice), and finally measure in the new basis and
outcome the value $a$ or $b$.

Consider quantum states that are Schmidt-diagonal in the
computational basis: \ba \ket{\psi}&=&\sum_{k=0}^{d-1}
c_k\,\ket{k\,k} \ea with $c_k\in\real$: on this family, one finds
\ba
P(a,b|x,y)&=&\frac{1}{d^2}\sum_{k,k'=0}^{d-1}\,c_kc_{k'}\,\times\nonumber\\&&
\cos\left[\left(\frac{2\pi}{d}\Delta+\phi_x+\theta_y\right)(k-k')\right]
\ea with $\Delta=a-b$. The only freedom left is the choice of the
four angles $\phi_{x}$ and $\theta_{y}$. The settings we are
interested in are defined by \ba \phi_0=0\,,\;
\phi_1=\frac{\pi}{d}\,;\; \theta_0=-\frac{\pi}{2d}\,,\;
\theta_1=\frac{\pi}{2d}\,.\ea With these settings, (\ref{slice})
holds.

For the case $d=3$, all the interesting states found to date are
of the form (\ref{psigamma}). For instance, $\gamma=1$ is the
maximally entangled state; the maximal violation is obtained for
$\gamma=\frac{\sqrt{11}-\sqrt{3}}{2}\approx 0.7923$ \cite{adgl02};
the largest Kullback-Leibler distance from the set of local
distributions is obtained for $\gamma\approx 0.6529$ \cite{agg05};
and we have shown above (\ref{secd3}) that the maximal amount of
secret key rate under one-way processing is found for
$\gamma\approx 0.9875$. For the states $\ket{\psi(\gamma)}$, the
$p_{\Delta}=P(a,\Delta-a|0,0)$ are: \ba
\begin{array}{lcl}p_0&=&
\frac{1}{3}\left(1+\frac{1+2\sqrt{3}\gamma}{2+\gamma^2}\right)\,,\\
p_1&=&\frac{1}{3}\left(1-\frac{2}{2+\gamma^2}\right)\,,\\
p_2&=&
\frac{1}{3}\left(1+\frac{1-2\sqrt{3}\gamma}{2+\gamma^2}\right)\,.\end{array}
\label{probsgamma}\ea

\section{Depolarization for arbitrary $d$}
\label{appdepolarize}

\subsection{The procedure}

Suppose Alice and Bob share initially an arbitrary no-signalling
probability distribution $P(a,b|x,y)$. The depolarization
procedure that brings $P$ in the slice defined by (\ref{slice}) is
very similar to the one described in Ref.~\cite{generic} for
$d=2$. It consists of two steps.

{\em Step 1.} Alice chooses $k\stackrel{R}{\in}\{0,...,d-1\}$ with
probability $\frac{1}{d}$ and communicates it to Bob on a public
channel. Both Alice and Bob perform \ba
\begin{array}{lcl} a&\longrightarrow a+k\\ b&\longrightarrow b+k
\end{array}\,.
\ea This implements $P\rightarrow P_1$ which is such that
$P_1(a,b|x,y)=\frac{1}{d}\sum_{k}P(a+k,b+k|x,y)$ and is
consequently a function only of $\Delta=a-b$.

{\em Step 2.} With probability $\frac{1}{4}$, Alice chooses one of
the following four procedures and asks Bob on the public channel
to act accordingly: \ba \begin{array}{lcl}
\mbox{Proc}_1&:&\begin{array}{ll}A:&\mbox{do nothing}\\
B:&\mbox{do nothing}\end{array}\\
\mbox{Proc}_2&:&\begin{array}{ll}A:&x\rightarrow \bar{x},\,a\rightarrow -a\\
B:&b\rightarrow -b+y\end{array}\\
\mbox{Proc}_3&:&\begin{array}{ll}A:&a\rightarrow -a-x\\
B:&y\rightarrow \bar{y},\,b\rightarrow -b\end{array}\\
\mbox{Proc}_4&:&\begin{array}{ll}A:&x\rightarrow \bar{x},\,a\rightarrow a+x\\
B:&y\rightarrow \bar{y},\,b\rightarrow b+\bar{y}\end{array}\\
\end{array}\ea where we have written $\bar{x}=1-x$. This implements
$P_1\longrightarrow P_2$ such that \ban
4P_2(a,b|0,0)&=&P_1(a,b|0,0)+P_1(-a,-b|0,1)\\
&&+P_1(-a,-b|1,0)+P_1(a,b+1|1,1)\\
4P_2(a,b|0,1)&=&P_1(a,b|0,1)+P_1(-a,-b+1|1,1)\\
&&+P_1(-a,-b|0,0)+P_1(a,b|1,0)\\ 4P_2(a,b|1,0)&=&P_1(a,b|1,0)+P_1(-a-1,-b|1,1)\\
&&+P_1(-a,-b|0,0)+P_1(a+1,b+1|0,1)\\ 4P_2(a,b|1,1)&=&P_1(a,b|1,1)+P_1(-a,-b+1|0,1)\\
&&+P_1(-a-1,-b|1,0)+P_1(a+1,b|0,0)\,. \ean Because of the symmetry
of $P_1$, this implies $P_2(a,b|0,0)=P_2(-a,-b|0,1)=
P_2(-a,-b|1,0)= P_2(a,b+1|1,1)$ which is nothing but the
definition of the slice (\ref{slice}).

\subsection{Examples}

We said in the main text that none of the extremal points of the
form $PR_{2,d'}$, with $d'<d$, is on the slice. Let's then
consider a realization of $PR_{2,d'}$, the one whose array is \ba
\hat{P}(d')&=& \frac{1}{d'}\,
\begin{array}{c|cc|cc|} A\setminus B& \mathbf{1} & \mathbf{0} &
\mathbf{1} & \mathbf{0}\\
\hline \mathbf{1} & \one_{d'} & \mathbf{0} &
\one_{d'} & \mathbf{0}\\
\mathbf{0} & \mathbf{0} & \mathbf{0} & \mathbf{0} &
\mathbf{0}\\
\hline \mathbf{1} & \one_{d'} & \mathbf{0} & U_{d'}
 & \mathbf{0}\\
\mathbf{0} & \mathbf{0} & \mathbf{0} & \mathbf{0} &
\mathbf{0}\\\hline\end{array}\label{reprdprime}\ea where boldface
numbers indicate arrays containing all ones or all zeros,
$\one_{d'}$ is the identity matrix of dimension $d'\times d'$, and
where $U_{d'}$ is the $d'\times d'$ matrix \ba
U_{d'}&=&\left(\begin{array}{cccccc}
0&1&0&0&\cdots&0\\ 0&0&1&0&\cdots&0\\
0&0&0&1&&0\\
\vdots&\vdots&&&\ddots&\\
0&0&0&0&\cdots& 1\\
1&0&0&0&\cdots&0
\end{array}\right)\,. \ea The arrays (\ref{cglmp}) and (\ref{reprdprime}) allow to compute
immediately the "scalar product" \ba
\left<I_d,\hat{P}(d')\right>&=&1-\frac{1}{d'}\ea generalizing the
results we gave in the main text for $d'=2$ and $d'=d$.

By following the steps of the depolarization protocol, one finds
that $\hat{P}(d')$ goes to the distribution in the slice which is
given by \ba \hat{P}(d')&\longrightarrow
&\hat{P}_2(d')\,\equiv\,\left\{
\begin{array}{lcl} p_0&=&1-\frac{1}{4d'}\\p_{d'}&=&
\frac{1}{4d'}\end{array}\right. \ea and obviously all the other
$p_f$ are zero. Using (\ref{pnl}), one can verify that
$\left<I_d,\hat{P}_2(d')\right>=1-\frac{1}{d'}$: the violation is
preserved. As we said in the main text, this is not peculiar to
this example, but is a general feature, as we show in the next
paragraph.

\subsection{Preservation of the violation of CGLMP}

We want to prove that this depolarization preserves the violation
of the CGLMP inequality, that is \ba \left<I_d,P_2\right>&=&
\left<I_d,P\right>\,. \ea The easiest way is to write down $I_d$
as it appeared in the original paper \cite{cglmp}, namely
$\tilde{I}_d\leq 2$ with \ban \tilde{I}_d &=&
\sum_{k=0}^{[d/2]-1}\left(1-\frac{2k}{d-1}\right)\,\Big\{\big[P(-k|0,0)+P(k|0,1)\nonumber\\
&&+P(k|1,0)+P(-k-1|1,1)\big]-
\big[P(k+1|0,0)\nonumber\\
&&+P(-k-1|0,1)+P(-k-1|1,0)+P(k|1,1)\big]\Big\} \ean where
$P(\Delta|x,y)\equiv P(a-b=\Delta|x,y)$. The link between
$\tilde{I}_d$ and our definition of $I_d$ is provided by \ba
I_d&=&\frac{d-1}{2d}\,\left(-2+\tilde{I}_d\right)\,. \ea Using the
expression of $\tilde{I}_d$, the proof is straightforward. In
fact, Step 1 keeps by definition all the $P(\Delta|x,y)$ constant,
while Step 2 keeps both sums in $[...]$ constant.

\section{Deterministic strategies that saturate CGLMP}
\label{apptechn}

We present here a more detailed study of the extremal points that
lie on the CGLMP facet, completing what has been written in
paragraph \ref{ssloc}.

Consider the array which represents the CGLMP inequality $I_d\leq
0$, eq.~(\ref{cglmp}); here, it is more convenient to look at it
as having $2d\times 2d$ entries \cite{note3}. Let $I[i,j]$ denote
an entry of this array. For the deterministic strategy
$\big\{a(0),a(1);b(0),b(1)\big\}$, the value of CGLMP is
simply \ba I_d&=& -2+\sum_{x,y=0}^1 I[a(x),b(y)]\nonumber\\
&=&-2\,+\,\delta[b(0)\geq a(0)] + \delta[a(0)\geq b(1)]\nonumber\\
&&+ \delta[a(1)\geq b(0)] - \delta[a(1)\geq b(1)] \label{iddet}\ea
where the $-2$ comes from the marginals of $a(0)$ and $b(0)$, and
where $\delta[C]$ is equal to 1 if condition $C$ is satisfied and
to 0 otherwise. The inequality is saturated by all the strategies
such that $I_d=0$.

Consider the points such that $b(0)=b(1)$: the last two conditions
become equal and the $\delta$'s compensate each other for all
$a(1)$, so the only way to saturate the inequality is to fulfill
both $b(0)\geq a(0)$ and $a(0)\geq b(1)=b(0)$; whence
$a(0)=b(0)=b(1)$ as announced in the main text. The proof of the
analog statement in the case $b(0)=b(1)-1$ is similarly done by
inspection. One first considers $b(0)<d-1$: in this case,
$b(1)>b(0)$, therefore the first two conditions cannot be both
fulfilled, whatever $a(0)$ is. One can then easily verify that
only the choice $a(1)=b(0)$ leads to a saturation of the
inequality. The last remaining case is $b(0)=d-1$, $b(1)=0$: it
can be read directly from the array, and leads to the same
conclusion.

So, we have proved the properties of sets ${\cal{L}}_0$ and
${\cal{L}}_1$, which consist of $d^2$ points each; we still have
to prove that the number of points on the facet is larger than
$D=4d(d-1)$. This is easily done by noticing the following: the
four "natural" relations associated to the CGLMP inequality, those
that are simultaneously fulfilled by $PR_{2,d}$, are: \ba
\begin{array}{lcl}
R_00&:&a(0)=b(0)\\ R_01&:&a(0)=b(1)\\ R_10&:&a(1)=b(0)\\
R_11&:&a(1)=b(1)-1\,.\end{array} \label{natural}\ea Because of the
specific pseudo-sifting of our cryptographic protocol, we grouped
them by pairs according to Alice's input. But from the standpoint
of the inequality, any pairwise grouping is equally meaningful. It
can indeed be easily verified using (\ref{iddet}) that all the
points that fulfill at least two among these relations saturate
the inequality. There are therefore $4d$ strategies that fulfill
three relations, and $6d(d-2)$ strategies that fulfill exactly two
relations. In conclusion, by looking only at the points that
fulfill at least two among the four relations (\ref{natural}), we
have already $6d^2-8d$ deterministic points on the CGLMP facet,
and this number is larger than $D$ for $d>2$. We note that the
list is exhaustive for $d=3$ (see Appendix \ref{apptechn3}), but
not in general. For instance, for $d\geq 5$, the strategy
$\big\{a(0)=4,a(1)=1;b(0)=5,b(1)=3\big\}$ fulfills none of the
relations (\ref{natural}), but achieves nevertheless $I_d=0$.

\section{Explicit analysis for $d=3$}
\label{apptechn3}

\subsection{Deterministic strategies on the facet}

We give here the explicit list of the 30 deterministic strategies
that saturate CGLMP. We note $r=0,1,2$.

The twelve strategies in ${\cal{L}}^3$ are \ba {\cal{L}}_0^3
&:&\left\{\begin{array}{lcl}
L_{3,1}^r &=& \{a(x)=r,b(y)=r\}\\
L_{3,2}^r &=& \{a(x)=r-x,b(y)=r\}
\end{array}\right. \\
{\cal{L}}_1^3 &:&\left\{\begin{array}{lcl}
L_{3,3}^r &=& \{a(x)=r,b(y)=r+y\}\\
L_{3,4}^r &=& \{a(x)=r-x,b(y)=r+y-1\}
\end{array}\right.\,.
\ea The six strategies in ${\cal{L}}^2$ are \ba {\cal{L}}_0^2
&:\;&
L_{2,1}^r \,=\, \{a(x)=r+x,b(y)=r\}\\
{\cal{L}}_1^2 &:\;& L_{2,2}^r \,=\, \{a(x)=r+x,b(y)=r+y+1\}\,.\ea
The twelve strategies outside ${\cal{L}}$ are: \ba
\begin{array}{lcl}
L_{e,1}^r&=& \{a(x)=r,b(y)=r-y\}\\
L_{e,2}^r&=& \{a(x)=r+x,b(y)=r-y\}\\
L_{e,3}^r&=& \{a(x)=r+x,b(y)=r-y+1\}\\
L_{e,4}^r&=& \{a(x)=r-x,b(y)=r-y+1\}\,.\end{array} \ea

As we said in the main text, the decomposition of $M_2$ given in
(\ref{decomp}) is only one possible decomposition, the one which
optimizes Eve's information on Bob's symbol. It can checked that
the general decomposition is defined by \ba
M_2&:\;&\begin{array}{lcl} p_{3,j}^r&=&0\,,\\
p_{2,1}^r&=&p_{2,2}^r\,\equiv \,p_2^r\mbox{ free,}\\
p_{e,1}^r&\mbox{free,}\\
p_{e,2}^r&=&\frac{1}{6}-(p_2^r+p_{e,1}^r)\,,\\
p_{e,3}^r&=&p_{e,1}^{r+1}\,,\\
p_{e,4}^r&=&\frac{1}{6}-(p_2^r+p_{e,1}^{r+1})\,.
\end{array} \label{decgen}\ea
There are thus six free parameters $\{p_2^r,p_{e,1}^r\}$,
constrained of course by the positivity of probabilities (in
particular, none of these parameters can exceed $\frac{1}{6}$). A
possible realization of $M_2$ is the equiprobable mixture of the
eighteen points which are not in ${\cal{L}}^3$. The choice leading
to (\ref{decomp}) is the equiprobable mixture of the six points in
${\cal{L}}^2$ ($p_2^r=\frac{1}{6}$, implying automatically
$p_{e,j}^r=0$).

\subsection{Alice-Bob-Eve correlations}

Having the explicit deterministic strategies, it is a matter of
patience to derive the Tables for the correlations Alice-Bob-Eve.
The result is given in Table \ref{tabled3gen}, in which we have
introduced the notations \ba f(p)=\frac{2p_1}{3}+2p_2p&\;,\;\; &
g(p)= \frac{1-p_0}{3}-2p_2p\,.\ea Note that, in each of the nine
cells, the sum of the probabilities does not depend on the
$p_2^r$, as it should: the decomposition of $M_2$ is known only to
Eve. Eve is obviously interested in maximizing the probability of
knowing both symbols, measured by $f(p)$; whence the choice
$p_2^r=\frac{1}{6}$ made in the main text.

\subsection{About non-local points that violate CGLMP}

Here, we want to list some non-local points other than $PR_{2,3}$,
that violate CGLMP, and study their relation with the slice
(\ref{slice}).

Consider first the non-local points equivalent to $PR_{2,2}\equiv
PR$. There are 24 such points in the no-signalling polytope: in
fact, there a three choices for the two outcomes [(0,1), (0,2) or
(1,2)] and for each choice there are eight PR-like points,
obtained as usual by relabelling inputs and/or outputs. By
inspection, it can be seen that $I_3>0$ (in fact, $I_3=\demi$) for
our representative (\ref{cglmp}) of $I_3$, is achieved only by
three PR-like points: those defined by \ba
\left\{\begin{array}{lcl} a=b &{\mbox{ if }}& xy=0\\
a\neq b &{\mbox{ if }}& xy=1\\
\end{array}\right. & \mbox{ for }
&(a,b)\in \left\{\begin{array}{l} (0,1)\\(0,2)\\(1,2)

\end{array}\right.\,.
\ea It is readily seen that no mixture of these three strategies
can belong to the slice (\ref{slice}): to obtain all the marginals
equal to $\frac{1}{3}$, the only possible mixture is the
equiprobable one. This one reads \ba M_{PR}&=& \frac{1}{3}\,
\begin{array}{c|cc|cc|} A\setminus B& 1&1& 1& 1\\\hline 1&
1&0&1&0\\ 1&0&1&0&1\\\hline 1&1&0&0&\demi\\ 1&0&1&\demi&0\\\hline
\end{array}\ea and is clearly not of the form (\ref{parray}). This
negative result is important for our study: had such a mixture
belonged to the slice, Eve would have sent these non-local points,
for which she would have gained some information (because in each
case one result is impossible).

Actually, there is a mixture of non-local points on the slice: it
is a mixture of other $PR_{2,3}$-like strategies, which optimize
the violation of different representatives of CGLMP, and violate
our representative by $I_3=\frac{1}{3}$. The strategies are those
in which $b-a$ is equal to $-xy$, $x(2-y)$, $y(2-x)$ and
$(x+y+1)\mbox{mod}2$ (note that this last one is indeed a
$PR_{2,3}$: the non-locality is embedded on the fact that the
r.h.s. is computed modulo 2, instead of modulo 3 as is the case
for the others). The equiprobable mixture of these four strategies
is the point $p_0=\frac{3}{4}$ and $p_1=0$ in the slice.
Obviously, Eve has no interest in sending these strategies instead
of the $PR_{2,3}$ which gives the maximal violation: in all cases,
she is going to learn nothing about the outcomes.

\end{appendix}

\section*{TABLES}

\begin{table}
\begin{tabular}{|c||c|c|c|c|}
$A\setminus B$ & $y=0, b=0$ & $y=0, b=1$ & $y=1, b=0$ & $y=1,
b=1$\\\hline\hline
$x=0,$ & $p_{NL}/2$ $(PR)$ &  & $p_{NL}/2$ $(PR)$&  \\
$a=0$& $p_1^0$ $(L_1^0)$ & & $p_1^0$ $(L_1^0)$ & \\
& $p_2^0$ $(L_2^0)$ & & $p_2^0$ $(L_2^0)$ & \\
& $p_3^0$ $(L_3^0)$ & $p_4^0$ $(L_4^0)$& $p_4^0$ $(L_4^0)$ &
$p_3^0$ $(L_3^0)$\\\hline
$x=0,$ & & $p_{NL}/2$ $(PR)$& & $p_{NL}/2$ $(PR)$\\
$a=1$&& $p_1^1$ $(L_1^1)$ & & $p_1^1$ $(L_1^1)$ \\
&& $p_2^1$ $(L_2^1)$ & & $p_2^1$ $(L_2^1)$ \\
&$p_4^1$ $(L_4^1)$& $p_3^1$ $(L_3^1)$ & $p_3^1$ $(L_3^1)$ &
$p_4^1$ $(L_4^1)$
\\\hline
$x=1,$ &$p_{NL}/2$ $(PR)$ & & & $p_{NL}/2$ $(PR)$ \\
$a=0$& $p_1^0$ $(L_1^0)$ & $p_2^1$ $(L_2^1)$ & $p_1^0$ $(L_1^0)$ & $p_2^1$ $(L_2^1)$ \\
& $p_3^0$ $(L_3^0)$ & &  & $p_3^0$ $(L_3^0)$ \\
& $p_4^1$ $(L_4^1)$ & &  & $p_4^1$ $(L_4^1)$
\\\hline
$x=1,$ & &$p_{NL}/2$ $(PR)$ &$p_{NL}/2$ $(PR)$ &\\
$a=1$& $p_2^0$ $(L_2^0)$ & $p_1^1$ $(L_1^1)$ & $p_2^0$ $(L_2^0)$ & $p_1^1$ $(L_1^1)$ \\
& & $p_3^1$ $(L_3^1)$ & $p_3^1$ $(L_3^1)$ &\\
& & $p_4^0$ $(L_4^0)$ & $p_4^0$ $(L_4^0)$ &
\end{tabular}
\caption{Table of the distribution Alice-Bob-Eve for the raw data.
The entries are the $P(a,b|x,y)$. In parentheses, we indicate
Eve's symbol.} \label{tableraw}
\end{table}

\begin{table}
\begin{tabular}{|c||c|c|}
$\mathbf{x=0}$ & $b=0$ & $b=1$ \\\hline\hline $a=0$ & $p_{NL}/2$
$(PR)$ &\\
&$p_1^0$ $(L_1^0)$ &\\
&$p_2^0$ $(L_2^0)$ &\\
&$p_3^0\xi_0$ $(L_3^0)$ & $p_3^0\xi_1$
$(L_3^0)$\\
&$p_4^0\xi_1$ $(L_4^0)$ & $p_4^0\xi_0$
$(L_4^0)$\\
\hline
$a=1$ & & $p_{NL}/2$ $(PR)$\\ &&$p_1^1$ $(L_1^1)$ \\
&&$p_2^1$ $(L_2^1)$ \\
&$p_3^1\xi_1$ $(L_3^1)$ & $p_3^1\xi_0$
$(L_3^1)$\\
&$p_4^1\xi_0$ $(L_4^1)$ & $p_4^1\xi_1$ $(L_4^1)$
\end{tabular}
\begin{tabular}{|c||c|c|}
$\mathbf{x=1}$ & $b=0$ & $b=1$ \\\hline\hline $a=0$ & $p_{NL}/2$
$(PR)$ &\\
&$p_1^0\xi_0$ $(L_1^0)$ & $p_1^0\xi_1$ $(L_1^0)$\\
& $p_2^1\xi_1$ $(L_2^1)$ & $p_2^1\xi_0$ $(L_2^1)$\\
&$p_3^0$ $(L_3^0)$ &\\
&$p_4^1$ $(L_4^1)$ &\\\hline
$a=1$ & & $p_{NL}/2$ $(PR)$\\
&$p_1^1\xi_1$ $(L_1^1)$ & $p_1^1\xi_0$
$(L_1^1)$\\
&$p_2^0\xi_0$ $(L_2^0)$ & $p_2^0\xi_1$
$(L_2^0)$\\
&&$p_1^3$ $(L_1^3)$ \\
&&$p_4^0$ $(L_4^0)$
\end{tabular}
\caption{Probability distributions Alice-Bob-Eve for the data
after the pseudo-sifting of the CHSH protocol, conditioned to the
knowledge of $x=0$ or $x=1$.} \label{tablesx}
\end{table}

\begin{table}
\begin{tabular}{|c||c|c|}
[isotropic] & $b=0$ & $b=1$ \\\hline\hline $a=0$ & $p_{NL}/2$
$(?,?)$ &\\
&$p_L/4$ $(0,0)$ &\\
&$p_L/8$ $(0,?)$ & $p_L/8$
$(0,?)$\\
\hline
$a=1$ & & $p_{NL}/2$ $(?,?)$\\ &&$p_L/4$ $(1,1)$ \\
&$p_L/8$ $(1,?)$ & $p_L/8$ $(1,?)$
\end{tabular}
\caption{Probability distribution Alice-Bob-Eve for the CHSH
protocol, in the case of isotropic distribution.} \label{tableiso}
\end{table}

\begin{table}
\begin{tabular}{|c||c|c|}
& $b=0$ & $b=1$ \\\hline\hline

$a=0$ & $\frac{p_{NL}}{2}(\bar{q}_A\bar{q}_B+{q}_A{q}_B)$ $(?,?)$
&
$\frac{p_{NL}}{2}({q}_A\bar{q}_B+\bar{q}_A{q}_B)$ $(?,?)$\\
&$\frac{p_L}{4}\bar{q}_A\bar{q}_B$ $(0,0)$ &
$\frac{p_L}{4}\bar{q}_A{q}_B$ $(0,0)$\\
&$\frac{p_L}{4}{q}_A{q}_B$ $(1,1)$ &
$\frac{p_L}{4}{q}_A\bar{q}_B$ $(1,1)$\\
&$\frac{p_L}{8}\bar{q}_A$ $(0,?)$ & $\frac{p_L}{8}\bar{q}_A$
$(0,?)$\\
&$\frac{p_L}{8}{q}_A$ $(1,?)$ & $\frac{p_L}{8}{q}_A$
$(1,?)$\\\hline

$a=1$ & $\frac{p_{NL}}{2}({q}_A\bar{q}_B+\bar{q}_A{q}_B)$ $(?,?)$&
$\frac{p_{NL}}{2}(\bar{q}_A\bar{q}_B+{q}_A{q}_B)$
$(?,?)$ \\
&$\frac{p_L}{4}{q}_A\bar{q}_B$ $(0,0)$ &
$\frac{p_L}{4}{q}_A{q}_B$ $(0,0)$\\
&$\frac{p_L}{4}\bar{q}_A{q}_B$ $(1,1)$ &
$\frac{p_L}{4}\bar{q}_A\bar{q}_B$ $(1,1)$\\
&$\frac{p_L}{8}{q}_A$ $(0,?)$ & $\frac{p_L}{8}{q}_A$
$(0,?)$\\
&$\frac{p_L}{8}\bar{q}_A$ $(1,?)$ & $\frac{p_L}{8}\bar{q}_A$
$(1,?)$
\end{tabular}
\caption{Probability distribution Alice-Bob-Eve, in the case of
isotropic distribution, after Alice's and Bob's pre-processing.}
\label{tableisopp}
\end{table}

\begin{table}
\begin{tabular}{|c||c|c|c|} & $b=0$ & $b=1$ & $b=2$ \\\hline\hline $a=0$ & $p_{NL}/3$
$(?,?)$ &&\\
&$\frac{p_L}{6}$ $(0,0)$ &&\\
&$\frac{p_L}{12}-\frac{p_2}{6}$ $(0,?_2)$ & $\frac{1-p_0}{6}$
$(0,?_2)$ & $\frac{1-p_0}{6}$
$(0,?_2)$\\
\hline
$a=1$ & & $p_{NL}/3$ $(?,?)$&\\ &&$\frac{p_L}{6}$ $(1,1)$ &\\
&$\frac{1-p_0}{6}$ $(1,?_2)$ & $\frac{p_L}{12}-\frac{p_2}{6}$
$(1,?_2)$ & $\frac{1-p_0}{6}$ $(1,?_2)$\\
\hline
$a=2$ & & & $p_{NL}/3$ $(?,?)$\\ &&&$\frac{p_L}{6}$ $(2,2)$ \\
&$\frac{1-p_0}{6}$ $(2,?_2)$  & $\frac{1-p_0}{6}$ $(2,?_2)$ &
$\frac{p_L}{12}-\frac{p_2}{6}$ $(2,?_2)$
\end{tabular}
\caption{Probability distribution Alice-Bob-Eve for $d=3$, after
pseudo-sifting, assuming decomposition (\ref{decomp}) for $M_2$.
We indicate by $?_2$ the case where Eve hesitates among two values
of Bob's symbol (instead of three).} \label{tabled3}
\end{table}

\begin{table}
\begin{tabular}{|c||c|c|c|}
\bf{x} & $b=0$ & $b=1$ & $b=2$ \\\hline\hline $a=0$ & $p_{NL}/3$
$(?,?)$ &&\\
&$f(p_2^{0-x})$ $(0,0)$ &&\\
&$g(p_2^{0-x})$ $(0,?_2)$ & $\frac{1-p_0}{6}$ $(0,?_2)$ &
$\frac{1-p_0}{6}$
$(0,?_2)$\\
\hline
$a=1$ & & $p_{NL}/3$ $(?,?)$&\\ &&$f(p_2^{1-x})$ $(1,1)$ &\\
&$\frac{1-p_0}{6}$ $(1,?_2)$ & $g(p_2^{1-x})$
$(1,?_2)$ & $\frac{1-p_0}{6}$ $(1,?_2)$\\
\hline
$a=2$ & & & $p_{NL}/3$ $(?,?)$\\ &&&$f(p_2^{2-x})$ $(2,2)$ \\
&$\frac{1-p_0}{6}$ $(2,?_2)$  & $\frac{1-p_0}{6}$ $(2,?_2)$ &
$g(p_2^{2-x})$ $(2,?_2)$
\end{tabular}
\caption{Probability distribution Alice-Bob-Eve after
pseudo-sifting for $d=3$ and Alice's setting $x$, for the general
decomposition (\ref{decgen}) of $M_2$.} \label{tabled3gen}
\end{table}

\end{multicols}


\begin{thebibliography}{99}
\bibitem{notec} We stress explicitly an important point. There are
several ways of guaranteeing that no information leakage has taken
place about $x$ and $y$. One of these ways is based on physics: if
the measurement-and-detection events of Alice and Bob are
space-like separated events, then relativity guarantees that no
signal has propagated from one location to the other. This is a
sufficient, but not a necessary condition for independence. Our
work only assumes that indeed there is no information leakage,
independently of the way this is achieved.
\bibitem{note0} If the polytope is embedded in a $D$-dimensional space, a facet is a hyperplane of dimension
$D-1$. To define such a hyperplane, one needs $D$ linearly
independent points --- just as one needs two different points to
define a line, three non-aligned points to define a plan etc.
\bibitem{noteintr}
This map looks conter-intuitive from Eve's point of view. Why
should she map $(0,?)$ and $(1,?)$, where she knows Alice's
variable, into $(?,?)$? Recall however that this map is only a
mathematical tool for the computation of an upper bound on the
secret-key rate.
\bibitem{note2} We could have used the same formalism for the case
$d=2$; we did not, because we found it preferable to tackle that
case without the introduction of a lot of notations. Anyway, all
that we write in Section \ref{secd} for general $d$ applies in
particular for $d=2$.
\bibitem{note3} The array (\ref{cglmp}) that defines $I_d$ has
exactly the same form, whether one considers it as having
$2(d-1)\times 2(d-1)$ entries, or in the "full" form which has
$2d\times 2d$ entries. It is in fact easy to verify that the all
the terms that are added cancel out.
\end{thebibliography}
\end{document}